\documentclass[11pt,a4paper]{article}

\usepackage[colorlinks=true, linkcolor=black!50!blue, urlcolor=blue, citecolor=blue, anchorcolor=blue]{hyperref}
\usepackage[font=small,labelfont=bf,margin=0mm,labelsep=period,tableposition=top]{caption}
\usepackage[a4paper,top=3cm,bottom=2.5cm,left=2.5cm,right=2.5cm,bindingoffset=0mm]{geometry}

\usepackage{graphicx,placeins}
\usepackage{float}
\usepackage{afterpage}
\usepackage{epsfig,cite}
\usepackage{amssymb}
\usepackage{amsmath}
\usepackage{dsfont}
\usepackage{multirow}
\usepackage{url}
\usepackage{xcolor}
\usepackage{float}
\usepackage{afterpage}

\usepackage{url}
\usepackage{hyperref}
\usepackage{booktabs}

\usepackage{tikz}
\usetikzlibrary{arrows}
\usepackage{tikz-3dplot}
\usepackage{enumitem}
\usepackage{hyperref}
\usepackage{cite}

\newcommand{\be}{\begin{equation}}
\newcommand{\ee}{\end{equation}}
\newcommand{\bea}{\begin{eqnarray}}
\newcommand{\eea}{\end{eqnarray}}
\newcommand{\bi}{\begin{itemize}}
\newcommand{\ei}{\end{itemize}}
\newcommand{\ben}{\begin{enumerate}}
\newcommand{\een}{\end{enumerate}}

\newcommand{\lp}{\left(}
\newcommand{\rp}{\right)}

\def\frac#1#2{{{#1}\over {#2}}}
\def\gsim{\mathrel{\rlap{\lower4pt\hbox{\hskip1pt$\sim$}}
    \raise1pt\hbox{$>$}}}         %greater than or approx. symbol
\def\lsim{\mathrel{\rlap{\lower4pt\hbox{\hskip1pt$\sim$}}
    \raise1pt\hbox{$<$}}}         %less than or approx. symbol

\newcommand{\draft}[1]{}

\def\beq{\begin{equation}}
\def\eeq{\end{equation}}

\def\lapprox{\lower .7ex\hbox{$\;\stackrel{\textstyle <}{\sim}\;$}}
\def\gapprox{\lower .7ex\hbox{$\;\stackrel{\textstyle >}{\sim}\;$}}

\usepackage{tabularx}
\newcolumntype{C}[1]{>{\centering\arraybackslash}p{#1}}

\begin{document}
\newgeometry{top=1.5cm,bottom=1.5cm,left=2.5cm,right=2.5cm,bindingoffset=0mm}
\vspace{-2.0cm}
\begin{flushright}
Nikhef/2019-042\\
\end{flushright}
\vspace{0.3cm}

\begin{center}
  {\Large \bf The Partonic Content of Nucleons and Nuclei}
\vspace{0.8cm}

Juan Rojo\\

 \vspace{0.4cm}
 
       {\it \small Department of Physics and Astronomy, VU University, NL-1081 HV Amsterdam, and\\[0.1cm]
Nikhef Theory Group, Science Park 105, 1098 XG Amsterdam, The Netherlands\\[0.1cm]
}

\vspace{1.0cm}

{\bf \large Abstract}

\end{center}

Deepening our knowledge of the partonic content of nucleons and nuclei represents a central endeavour of
modern high-energy and nuclear physics, with ramifications in related disciplines
such as astroparticle physics.
There are two main scientific drivers motivating these investigations of the partonic structure
of hadrons.
On the one hand, addressing fundamental open issues in our understanding
of the strong interaction such as the origin of the
nucleon mass, spin, and transverse structure; the presence of heavy quarks in the nucleon
wave function; and the possible onset of novel gluon-dominated dynamical regimes.
On the other hand, pinning down with the highest possible precision the
substructure of nucleons and nuclei is a central component for theoretical
predictions in a wide range of experiments, from proton and heavy ion
collisions at the Large Hadron Collider to ultra-high energy neutrino interactions
at neutrino telescopes.
This Article presents a succinct non-technical overview of our
modern understanding of the quark, gluon,
and photon substructure of nucleons and nuclei, focusing on recent trends and results 
and discussing future perspectives for the field.\\[0.2cm]

\begin{center}

  {\it \large Invited Review Article to appear in the Oxford Research Encyclopedia of Physics}\\[0.5cm]

  
\end{center}  

\clearpage

% Length and scope. Your article should be 6,000-10,000 words in length, including Summary, Keywords, Main Essay, and References. Limit the use of jargon and abbreviations and define uncommon technical terms.

\tableofcontents

\section{Summary}

% The article summary should be a brief synopsis of the topic, not of the article itself. The summary should be roughly equivalent to a definition, one or two paragraphs in length. Unlike a traditional “abstract,” the summary should be able to stand on its own as a useful piece of content without reference to a larger article. Please do not include citations in the summary. The summary will publish right away and serve as a preview for the full article. After the article is published, the summary will appear at the beginning. If you would like to make changes to your summary when you submit the final article, please include a revised copy.

Protons and neutrons, collectively known as nucleons, represent together with
electrons the fundamental building blocks of matter and dominate the
overall mass of the visible Universe.
Nucleons, as well as all other hadrons, are characterised
by a rich internal substructure, being composed of elementary
particles, quarks and gluons,
collectively known as {\it partons}.
These partons are tightly held together within nucleons
thanks to the properties of the quantum field theory
of the strong interaction: Quantum Chromodynamics (QCD).

The study of the partonic content of nucleons is one of the
central endeavors of
modern high-energy and nuclear physics.
It is motivated by a number of fundamental
open questions in our understanding of the strong interaction
such as the origin of the nucleon mass and spin,
the three-dimensional profiling of hadron substructure,
the role
of heavy quarks in hadronic wave functions, and the potential onset of
novel gluon-dominated dynamical regimes.
In addition, deepening our knowledge of this partonic content
of nucleons is of paramount importance for a wide range of theoretical
predictions in high-energy process such as proton collisions
at the Large Hadron Collider (LHC), heavy ion collisions
at the Relativistic Heavy Ion Collider (RHIC), and high-energy
neutrino interactions
such as neutrino telescopes.
Furthermore, the quark and gluon substructure of nucleons is modified
once the latter become bound within heavy nuclei, leading to a remarkable
pattern of nuclear effects whose study opens a novel window
to the inner workings of the strong force in the nuclear environment.

The detailed investigation of the partonic content of nucleons and nuclei
is however a challenging task.
It requires the careful
combination of state-of-the-art theoretical calculations and the widest
possible range of experimental measurements by means of a statistically
robust and efficient fitting methodology.
This framework, the so-called {\it global QCD analysis} of the nucleon structure, has
experienced
remarkable progress in the recent years.
Some important milestones include the assessment of the  constraints
on the proton structure provided by LHC measurements;
the implementation of machine learning algorithms that
speed up dramatically the analysis while minimising procedural biases;
the exploitation of precision-frontier
calculations in both the strong and electroweak interactions;
the deployment of tailored advanced computational tools such
as fast interfaces;
novel methods to extract with high precision the photon content of the proton;
an improved
characterisation of the role of the initial state of heavy ion collisions
for Quark-Gluon Plasma studies;
and the development of novel methods
to estimate and propagate relevant sources of uncertainties to the final theory
predictions.

Despite all these achievements,
a long road still lies ahead with pressing open questions,
both of a theoretical and experimental nature, which are directly
relevant for the full physics exploitation of current and future
facilities.
Some of these include puzzling results concerning
the strange quark content of protons; how gluons behave
at very small and very large momentum fractions; the impact of
lattice QCD simulations; the specific
pattern of  modifications in the partonic structure of heavy nuclei induced
by nuclear effects; and the interplay between Standard
Model measurements and searches for New Physics at the high-energy frontier.

% Main Essay
%Each article should present an overview of the full scope of a topic, its animating factors, and its developmental arc. Discuss the observational, theoretical, and experimental techniques used on the phenomena of focus. Article structure can be devised in this fashion:
%Introductory Paragraphs (400 – 500 words)
% Define the topic you will cover and why.
% Outline the areas of science that inform your work. Note how this work fits in the larger context
%of physics.

\section{Introduction}
\label{sec:introduction}

Elementary particles such as leptons, {\it e.g.} electrons and neutrinos, do not have any known substructure.
As opposed to those, hadrons
(particles that experience the strong nuclear force) turn out not to be elementary but rather
bound states
composed of quarks and gluons, collectively known as partons.
These partons are tightly held together within hadrons by virtue
of the properties of the quantum
theory of the strong interaction, QCD.
Indeed, the mathematical structure of QCD implies that color-charged particles such as
quarks
(color being the analog of the electric charge in the strong interaction)
cannot exist in isolation and need to be confined within hadrons.

In addition to the light quarks (up and down) and gluons, hadrons contain also heavier
quarks (strange and charm quarks in particular) as well as a photon component.
Furthermore, the behaviour of partons is modified once protons and neutrons
(denoted as nucleons)
become themselves the building blocks of heavy nuclei, reflecting a
rich pattern of nuclear dynamics.
Several of the most
important properties related to the partonic constituents of hadrons, including
their longitudinal and transverse momentum and spin distributions, 
are determined by the poorly
understood non-perturbative regime of the strong force.
In this regime, first principle calculations are challenging
and most partonic properties need to be extracted
from experimental data by means of a global QCD analysis.

\begin{figure}[h]
  \begin{center}
    \includegraphics[width=0.99\textwidth]{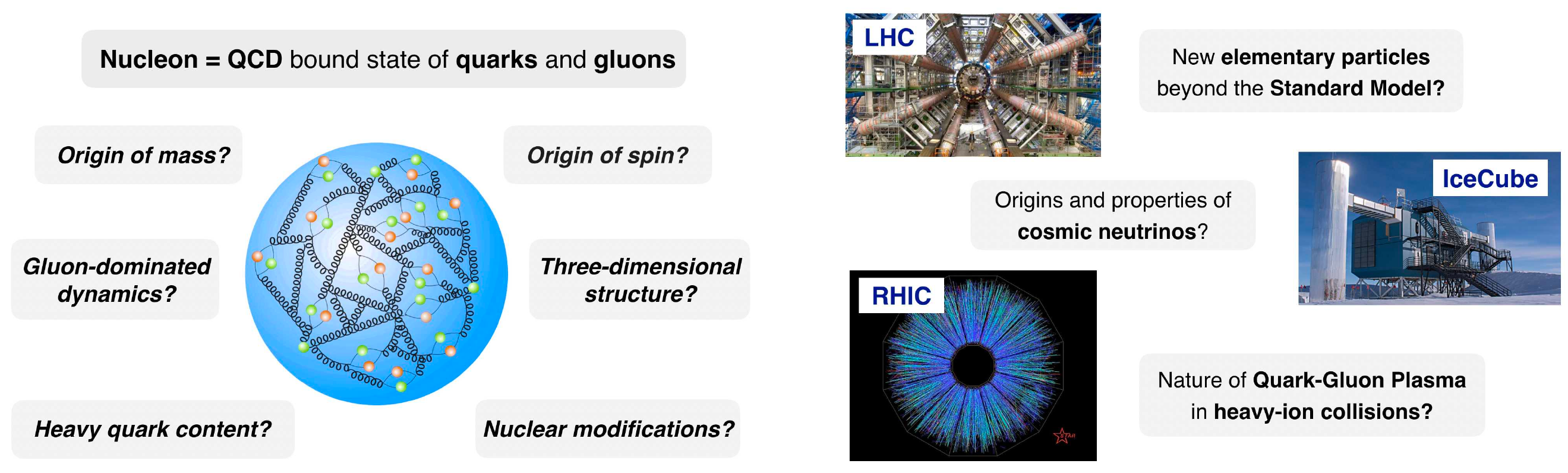}
    \caption{\small The investigation of the partonic content of nucleons and
      nuclei has a two-fold motivation: to address fundamental open issues
      in the strong interaction (left) and to provide precise theoretical
      predictions for experiments such as the LHC, IceCube, or RHIC (right).
        \label{fig:review-intro} }
  \end{center}
\end{figure}

Understanding the partonic content of hadrons and heavy nuclei plays a crucial role
in modern particle, nuclear, and astroparticle physics from a two-fold perspective,
summarised in Fig.~\ref{fig:review-intro}.
On the one hand, these partonic properties offer a unique window to address fundamental questions about
the inner workings of
the strong interaction and hadron structure, such as what is the origin of the nucleon mass
and spin, how is the motion of quarks and gluons modified inside heavy nuclei, or what determines
the onset of new states of matter where gluons dominate.
On the other hand, a precise quantitative description
of the partonic substructure of nucleons
is an essential
input for any theoretical predictions for a variety of experiments, from proton-proton scattering at the high-energy frontier
at the Large Hadron Collider (LHC)
to the collisions between heavy ions and to the interpretation
of the data provided by neutrino and cosmic ray telescopes.
In this respect, the determination of the partonic content of hadrons and nuclei offers
a bridge between different areas of modern physics and of related disciplines
such as advanced statistics, data analysis, and machine learning.

%%%%%%%%%%%%%%%%%%%%%%%%%%%%%%%%%%%%%%%%%%%%%%%%%%%%%%%%%%%%%%%%%%%
%%%%%%%%%%%%%%%%%%%%%%%%%%%%%%%%%%%%%%%%%%%%%%%%%%%%%%%%%%%%%%%%%%%%%%

This Article presents a succinct introduction to our
modern understanding of the quark, gluon,
and photon content of nucleons and nuclei, focusing on recent results and trends.
It also outlines and highlights possible future directions for the field of
proton structure studies.
The Article is deliberately non-technical, and for a detailed
discussions on the various topics covered here and an extensive survey of the related
literature the reader is encouraged to consult recent
topical reviews~\cite{Forte:2013wc,Rojo:2015acz,Gao:2017yyd,Kovarik:2019xvh}.

%Part One (2500 – 3500 words)
% Chart our understanding of the topic as it has developed over time: consider when and how the topic appeared and then took on its current form.
% Provide balanced coverage of the context, the controversies, and the debates that have informed and helped to form the topic, and that animate it now.
% Discuss foundational and notable discoveries or advances and those who made them within their context and current perspectives; include biographical details as needed.

\section{From the quark model to the Higgs boson}

Pushing forward our understanding of the partonic structure of nucleons and nuclei has been
at the forefront of fundamental research in particle and nuclear physics for more than five decades.
The story of quarks and gluons starts around the early 1930s.
By then, both the proton and neutron had been discovered, and thus the structure of atomic nuclei
could be explained.
At that time protons and neutrons were assume to be point-like particles without further
substructure, much in the same way as electrons.
For a long time, protons and neutrons were the only known {\it hadrons}, that is,
particles that
interacted via the strong nuclear force, responsible in particular for keeping
the atomic nucleus bound together.

The situation changed dramatically in the late 1940s with the discovery of neutral
and charged pions in cosmic ray experiments
- those were the main toolbox of particle physicists  before  accelerators became powerful
enough.
The discovery of the pions was followed by discoveries of a plethora of other strongly interacting particles:
kaons, rhos, lambdas, and omegas, just to name a few.
Each of these new hadrons was
characterised by different masses, electric charges, and spins (the intrinsic angular momenta
of quantum particles).
Physicists were confused and asked themselves how it was possible to establish some order in this chaos.
In other words, what where the underlying laws of Nature that determined the properties
of the large number of hadrons observed?
Furthermore, these new particles did not seem to play any obvious role in the properties
of everyday matter, with the possible exception of the pion that was thought to mediate
the strong interaction.
In addition, already since the 1930s the measurements of the anomalous magnetic moment
of the proton and the neutron suggested that these were not point-like particles
as {\it e.g.} electrons.

The first breakthrough towards clarifying this confusing
situation took place in the 1960s,
when  Gell-Mann and Zweig separately realised~\cite{GellMann:1964nj,Zweig:1964jf} that
the observed regularities in the hadron spectrum could be explained by assuming that
they were not fundamental particles, but instead bound states composed of new
 hypothetical  particles named {\it quarks}.
These newly proposed elementary particles were characterised by being point-like, having a fractional charge
of either $\pm2/3$ or $\pm 1/3$ in units of the electron charge, and half-integer spin.
Furthermore, this {\it quark model} assumed that quarks existed in three different types
or ``flavours'': the up, down, and strange quarks, which differed in their masses
and in some cases also in their electric charges.
It was then possible to show that by combining these three
quarks in various ways one could reproduce
the quantum numbers of most observed {\it baryons} (half-integer spin hadrons, assumed to be composed
by three quarks) and
{\it mesons} (integer-spin hadrons, composed of a quark-antiquark pair).

However, this quark model was purely phenomenological, and did not provide in particular
a suitable mechanism to explain
why quarks were tightly bound together within the hadrons.
Furthermore, while the quark model was successful in describing
hadron spectroscopy, it was unable to provide predictions for the observed
behaviour of strongly interacting particles in high-energy scattering
experiments.
For these reasons, physicists were for some time skeptical of the concept of quarks: while certainly
a useful mathematical framework
to describe hadron structure, their actual existence as real elementary particles
was far from widely accepted.
Such a skeptical position was further strengthened by the fact that free (isolated)
quarks had never been observed: even worse, no particles with fractional charge had ever been found
in Nature.

The road towards the acceptance of quarks as {\it bona fide} elementary particles was paved
by two momentous discoveries that took place in the late 1960s and early 1970s.
The first of those arose in a series of experiments at the Stanford Linear Accelerator Center (SLAC),
where energetic electrons where used as projectiles and fired at protons and neutrons
in atomic nuclei.
By investigating the properties of the deflected electrons, physicists could probe for the first time the
possible substructure of protons, in the same way as Rutherford's experiments at the beginning
of the century had stablished the existence of the atomic nucleus.

The SLAC measurements of this process~\cite{Bloom:1969kc}, known as deep-inelastic scattering (DIS), were
consistent
with a model for the
proton composed of point-like, non-interacting constituents of fractional electric charge,
thereby providing the long-sought experimental evidence for the existence of quarks.
Since those groundbreaking experiments, lepton-hadron DIS experiments such
as the SLAC ones have provided
the backbone for our investigations of the partonic structure of nuclei.
However, from the theoretical point of view there were still important hurdles
to be overcome before the quark model could be adopted universally.
In particular,  an explanation was needed for the fact that quarks appeared to
be free within the nucleon, while strong interaction scattering processes were characterised
in general by higher rates than the electromagnetic and weak ones.
The latter properties implied
the presence of a large coupling that determined the strength of the interaction,
seemingly inconsistent with the SLAC measurements.
How was it possible that the same interaction appeared to be either strong or weak
depending on the specific process?

The second revolution that lead to the modern understanding of the quark substructure of hadrons
was provided by the realisation in 1973 by  Gross, Politzer, and Wilczek
that the strong nuclear
force could be described in the framework of a renormalizable quantum field theory~\cite{Gross:1973id,Politzer:1973fx} in the same way as the electromagnetic
force was described by Quantum Electrodynamics (QED).
This new theory, called Quantum Chromodynamics (QCD),
was formulated in terms of quark fields interacting
 via the exchange of force mediators called {\it gluons}
that transmitted color charges, therefore playing
an analogous role as the photon in electromagnetism.
A crucial prediction of this new theory was that the strength of the interaction decreased for small
distances, or alternatively for high energies, thus explaining why in high-energy events quarks appeared
to be essentially free particles while at low energies, where the coupling strength increases,
quarks were bound within the hadrons.

Other predictions of QCD were subsequently confirmed, in particular evidence for the
existence of gluons
was obtained in electron-positron collisions in the late 1970s.
In these events,
the gluon appeared as a third stream of collimated hadrons
(known as a jet) in addition to the two associated with a quark-antiquark
pair~\cite{Brandelik:1979bd}.
In the following years, the new theory of the strong interaction was also
extended with additional heavier quarks, up to a total of 6 different
flavours, with the discovery of the charm (1972), bottom (1977), and top (1995) quarks.
The charm and bottom quarks have masses around 1.5 and 5 times the proton mass respectively,
and play an important role in the structure of heavy mesons and baryons.
The top quark, of the other hand, is too heavy (around 185 times the proton mass)
and decays before it can form bound states with other quarks.

The combination of the predictive power of the
quark model, the formulation of Quantum Chromodynamics as the quantum
field theory of the strong nuclear force, and
the results of high-energy experiments such as SLAC's deep-inelastic
scattering confirmed beyond reasonable doubt that hadrons were bound states composed
by quarks
and gluons, tightly held together by the non-perturbative phenomena that dominate the strong
nuclear force at low energies.
It became customary to collectively denote quarks and gluons, as well as any
eventual further component of hadrons, as {\it partons}.
Early on, it was  realised that the investigation of the partonic structure of
nucleons\footnote{The same considerations hold other hadrons.
  This Article
  concentrates on the partonic structure of protons, for which
  there is far more experimental information is available in comparison to other hadrons.}
was going to be a challenging endeavor.
Indeed, crucial properties of quarks and gluons, such as their
contribution to the total momentum and spin of the parent proton,
are determined by low-energy non-perturbative QCD dynamics.
This non-perturbative character implied that first principle calculations of the partonic
structure of nucleons could not be carried out within the framework of perturbation
theory, where the use of Feynman diagrams had lead to striking successes
in the case of weakly coupled theories such as Quantum Electrodynamics.

While perturbative QCD could not be applied to determine important
properties of hadron structure due to the dominance of the strong
coupling regime,  it was  realised  around the late 70s that
pQCD nevertheless could be used to impose important constraints on these properties.
In the same way as the coupling constant of the strong interaction
$\alpha_s(Q)$ depends on the momentum transfer involved in the process $Q$,
also partonic properties exhibit a similar energy dependence.
In particular, pQCD allowed to evaluate the dependence with the energy
of the momentum and spin fractions carried by quarks and gluons in the proton.
In other words, if one manages to determine that gluons carry a given fraction of the proton's momentum
at one specific energy, one can then predict how this fraction will be modified
for any other energy.
These so-called {\it evolution equations} were formulated by Dokshitzer, Gribov, Lipatov,
Altarelli and Parisi~\cite{gl,ap,dok} (DGLAP) and
have played an instrumental role to validate QCD as the correct theory
of the strong force at high energies.
Their predictions were spectacularly confirmed by the measurement of hard-scattering
observables at the HERA  lepton-proton collider~\cite{Klein:2008di} in the 90s.

Already during the dawn of the studies of the partonic
structure of hadrons, it became clear that
in order to make progress two main strategies could be pursued.
The first one is based on a first-principles
approach, whereby QCD is discretised on a space-time lattice
allowing different non-perturbative quantities to be evaluated by means
of powerful but very intensive computer simulations.
Until recently, the investigation of partonic properties using lattice QCD
was however rather limited due to a number of both conceptual
and computational bottlenecks.
This approach has experienced significant progress
in the recent years, as reviewed in the 2017 community White Paper~\cite{Lin:2017snn},
and is able now to provide
valuable information on the properties of quark and gluons within hadrons.

The second strategy exploits a central property of QCD known as {\it factorisation}, whereby
the total interaction cross-section $\sigma$, which is proportional to the number
of events of such type that will take place in a given period of time,
 can be separated into two independent
contributions for processes involving
hadrons in  the initial state of
the collision.\footnote{Similar factorised expressions apply for processes
that feature hadrons in the final state of the collision.}
The first one is the short-distance hard-scattering
partonic cross-section,
calculable using Feynman diagrams in the same way as in QED.
The second contribution encodes the information on the proton
structure determined by
long-distance, non-perturbative dynamics into objects
called {\it parton distribution functions} (PDFs).
The experimental measurements of such cross-sections can then be used, by virtue
of their factorisable properties, to extract the PDFs and thus achieve 
powerful insight into the partonic content and properties of protons.

This latter approach is usually
known as a {\it global QCD analysis}, and it
has been successfully deployed
in the last three decades to understand in increasing
detail the partonic structure
of nucleons and nuclei, as will be explained in the rest of this Article.
There exist various different types of PDFs, for example one can consider either spin-dependent
or spin-independent PDFs, collinear-integrated, or transverse momentum dependent PDFs.
This Article focuses on the collinear unpolarised
PDFs of nucleons and nuclei, which are, in the majority of the cases, the relevant
quantities for the interpretation
of the results of high-energy experiments such as the
proton-proton collisions at the Large Hadron Collider (LHC).
The interested reader can find further information about spin-dependent
PDFs in~\cite{Aidala:2012mv,Nocera:2014gqa} and about transverse-momentum dependent PDFs
in~\cite{Angeles-Martinez:2015sea} and references
therein.

In the following the following notation will be adopted
for the collinear unpolarised parton distribution functions of the proton:
\be
\label{eq:pdfdef}
\{\, f_i(x,Q^2) \,\} \, , \qquad i= 1, \ldots, n_f \, ,
\ee
where $x$ stands for the fraction of the nucleon's  momentum carried by the
$i$-th parton (often denoted as the Bjorken variable) and
$Q$ denotes the energy scale (which also
corresponds to the inverse of the resolution length and the momentum
transferred in the scattering process) at which the nucleon is being probed.
In Eq.~(\ref{eq:pdfdef}), the index $i$ indicates the parton flavour, and
$n_f$ denotes the number of active partons at the scale $Q$.
A quark is typically considered
a massless parton when the momentum transfers involved are larger than its
mass, $Q \gsim m_q$; otherwise it is assumed to be a massive state that does
not contribute to the partonic substructure of the proton.
For instance, at $Q=10$ GeV one would have $n_f=11$, given that
there are five quark PDFs
(up, down, strange, charm, and bottom) and the corresponding antiquark counterparts,
supplemented by the gluon PDF.

At leading order (LO) in the QCD perturbative expansion in the strong coupling
$\alpha_s$, the PDFs can be interpreted as probability densities.
This property implies that $g(x,Q=10\,{\rm GeV})\,dx$ would correspond
to the number of gluons
in the proton at a scale of  $Q=10$ GeV that carry a momentum fraction between $x$ and $x+dx$,
and likewise for other quark flavour combinations.
However, it is important to emphasize that
this naive probabilistic interpretation is lost when higher order effects in QCD are taken into
account.
In particular, it can be shown that PDFs become dependent on the specific
manner in which higher-order perturbative
corrections are evaluated, and thus cannot be associated to probabilities
nor to any specific physical observable.

\begin{figure}[h]
  \begin{center}
    \includegraphics[width=0.44\textwidth,angle=-90]{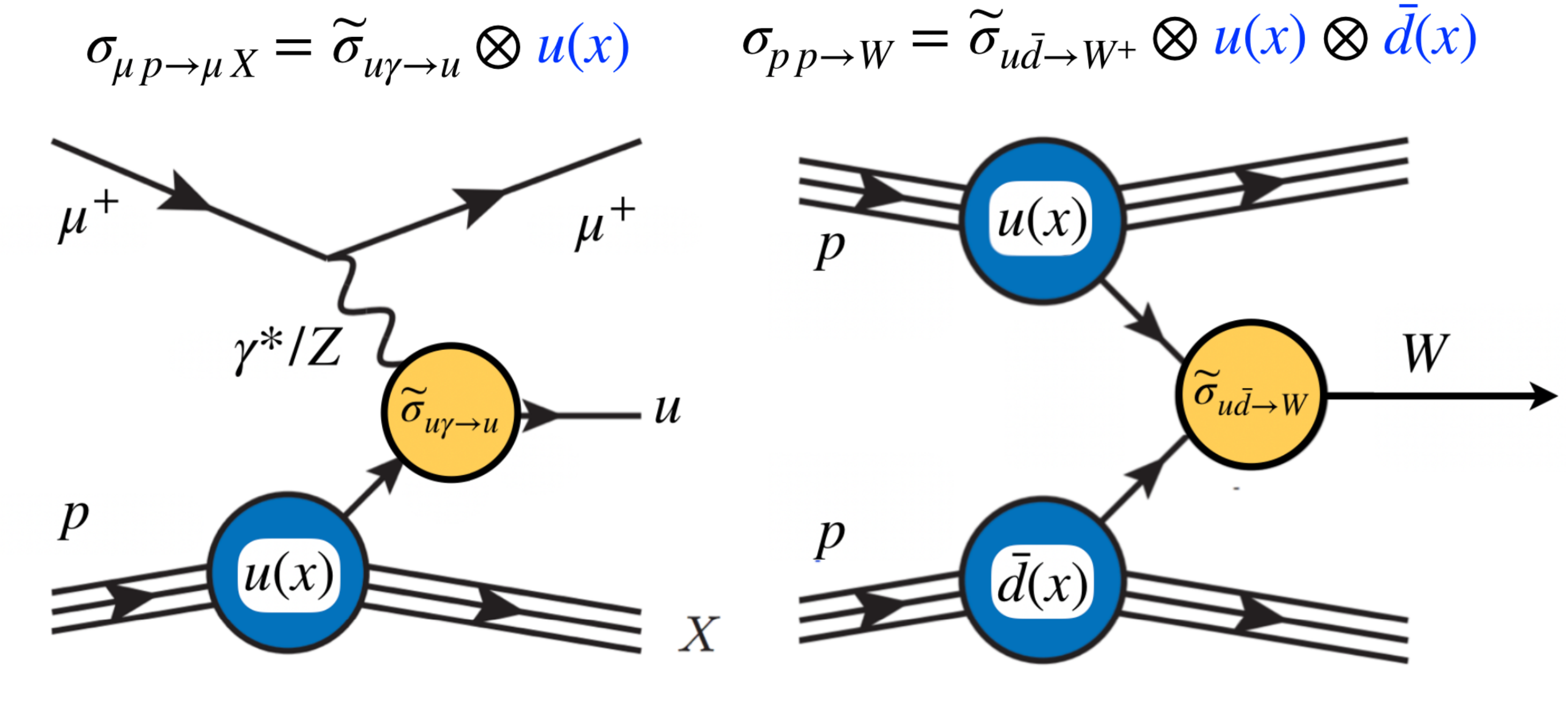}
    \caption{\small
      Left: the Feynman diagram associated
      with the deep-inelastic scattering process, where an energetic muon scatters
      off one of the quarks in the proton via the exchange of either a virtual
      photon or a $Z$ boson.
      Right: the corresponding diagram for $W$ boson production in proton-proton
      collisions, known as the Drell-Yan process.
      By virtue of the 
      QCD factorisation theorems and the PDF universality properties,
      it is possible to extract the proton PDFs
      from lepton-proton collisions (left) and then use the same PDFs
      to predict cross-sections in proton-proton collisions (right).
        \label{fig:factorisation} }
  \end{center}
\end{figure}

As mentioned before,
the dependence of the PDFs on the momentum fraction $x$ is determined by
low-energy non-perturbative
QCD dynamics, a regime where the strong coupling becomes large
and  perturbative calculations are not reliable.
Therefore, PDFs need to be extracted from experimental data as follows.
Consider the processes depicted in the left panel of Fig.~\ref{fig:factorisation},
where a charged lepton such as a muon scatters at high energy off a proton.
In such a process, which is of the same family of deep-inelastic scattering measurements
as those used in the pioneering SLAC experiments,\footnote{Note
  that the SLAC experiments used an electron beam instead, but the argument is the same
for muons.} the muon emits a virtual gauge
boson (either a photon or a $Z$ boson) which then interacts with one of the quarks in the proton,
in the example here an up quark.
The cross-section for this processes can be
schematically expressed by
\be
\label{eq:DIS1}
\sigma_{\mu\,p \to \mu\,X}(Q) = \widetilde{\sigma}_{\gamma^*\, u \to u}(Q) \otimes f_u(x,Q) \, ,
\ee
where $Q$ is related to the momentum transfer between
the muon and the proton 
and $\widetilde{\sigma}_{\gamma^*\, u \to u}$ is the hard partonic cross-section
for photon-quark scattering
that can be computed in perturbation theory using Feynman diagrams.
From Eq.~(\ref{eq:DIS1}) one can observe that if the hadronic cross-section
$\sigma_{\mu\,p \to \mu\,X}$ is measured and
the partonic one $\widetilde{\sigma}_{\gamma^*\, u \to u}$ can be evaluated
using perturbative QCD calculations, then one can extract the PDF of the up quark
$u(x,Q)$ from the data.

%%%%%%%%%%%%%%%%%%%%%%%%%%%%%%%%%%%%%%%%%%%%%%%%%%%%%%%%%%%%%%%%%%%
%%%%%%%%%%%%%%%%%%%%%%%%%%%%%%%%%%%%%%%%%%%%%%%%%%%%%%%%%%%%%%%%%%%%%%

At this point it is important to recall that, as mentioned
in Sect.~\ref{sec:introduction},
while the dependence of the PDFs on the momentum fraction $x$ (the Bjorken variable)
is indeed non-perturbative, their dependence on the momentum transfer $Q$ can be evaluated
in perturbation theory.
This can be achieved using the DGLAP evolution equations,
which take the schematic form
\be
Q^2 \frac{\partial}{\partial Q^2}f_i(x,Q^2) = \sum_{j=1}^{n_f} P_{ij}(x,\alpha_s(Q^2))\otimes
f_j(x,Q^2) \, , \qquad i=1,\ldots \, n_f \, ,
\ee
where $n_f$ indicates the number of active partons that participate in the evolution,
and $P_{ij}$ are perturbative kernels (the splitting functions) currently
known up to $\mathcal{O}\lp \alpha_s^4\rp$.
Thanks to these DGLAP equations, once 
the partonic structure of the nucleon
has been determined at some low scale, say $Q_0\simeq 1$ GeV,
one can evaluate the behaviour of the PDFs for any other scale $Q > Q_0$.
The splitting functions $P_{ij}(x,\alpha_s(Q^2)$ model the probability that
one of the partons in the proton ({\it e.g.} a quark
or a gluon) will radiate another parton in the collinear
approximation, where the two final-state particles move roughly in the same
direction).
Such a emission of a collinear quark or gluon will then modify the momentum fraction
$x$ carried by the original parton, with a probability
that depends on the scale $Q$ at which PDFs are being probed.

In addition to the constraints imposed by DGLAP evolution, there exist additional
theory considerations that restrict the behaviour of selected PDF combinations.
In particular, energy-momentum conservation imposes the momentum sum rule,
\be
\int_0^1 dx\,x\,\lp \sum_{i=1}^{n_q} \lp f_{q_i}(x,Q^2) + f_{\bar{q_i}}(x,Q^2)\rp + f_g(x,Q^2)\rp =1 \, ,
\ee
with $n_q$ the number of active quarks with masses $m_{q_i} < Q$,
which translates the property that the sum of the energies carried by all partons should
add up
to the total proton energy, while quark flavour number conservation leads to the
quark number sum rules,
\be
\label{eq:valenceSumRules}
\int_0^1 dx\,\lp u(x,Q^2)-\bar{u}(x,Q^2)\rp = 2 \, ,\qquad
\int_0^1 dx\,\lp d(x,Q^2)-\bar{d}(x,Q^2)\rp = 1 \, ,
\ee
reflecting the fact that the proton's wave function contains two  up quarks and one down quark.
Note that these sum rules should hold for any value of the momentum
transfer $Q$, and indeed one
can verify that DGLAP evolution ensures that if they are satisfied at some scale $Q_0$,
they will also be satisfied for all other scales.

\begin{figure}[h]
  \begin{center}
    \includegraphics[width=0.65\textwidth,angle=-90]{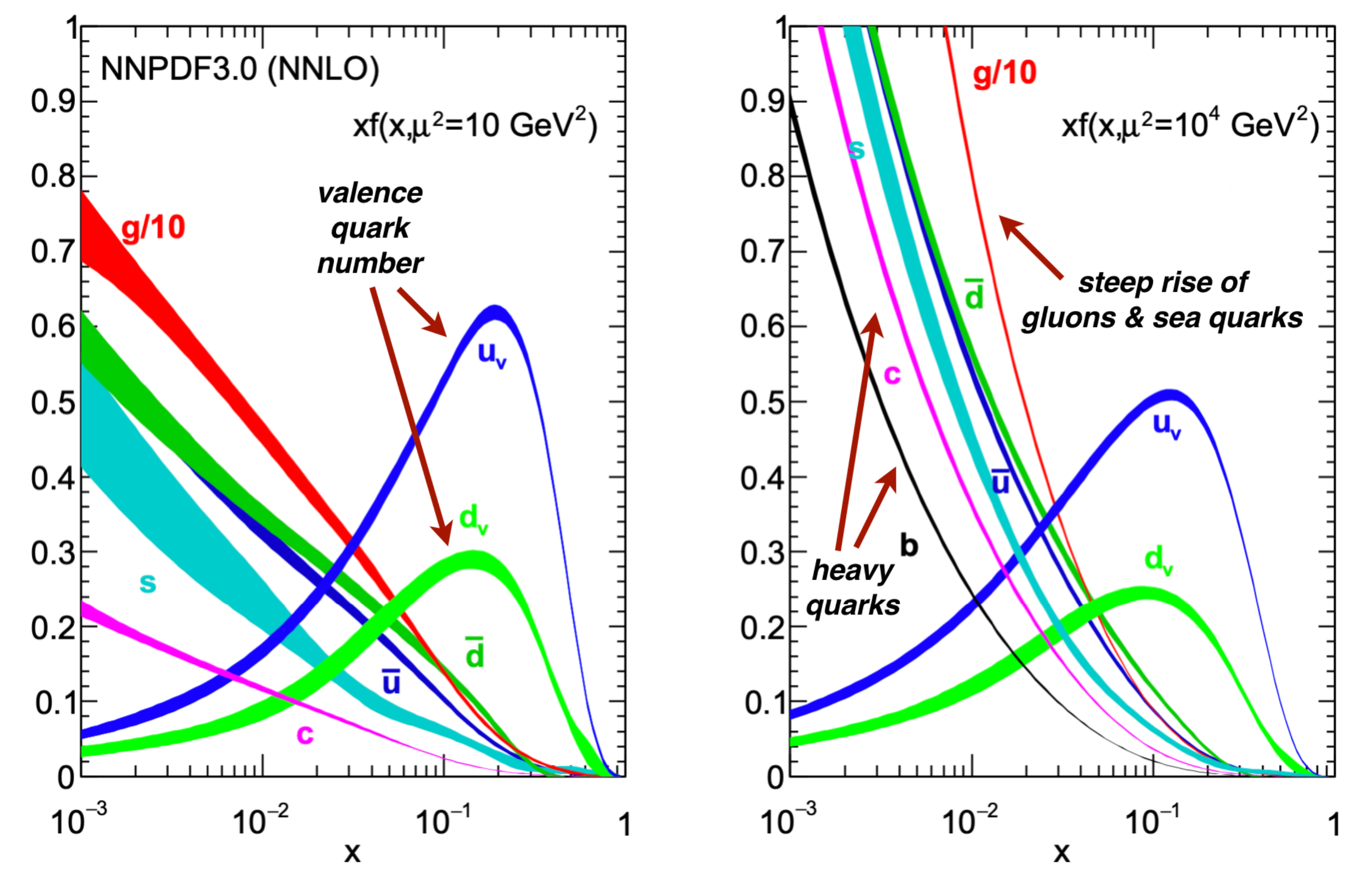}
    \caption{\small The results of the NNPDF3.0 NNLO global analysis
      for  $Q^2=10$ GeV$^2$ (left)
and for $Q^2=10^4$ GeV$^2$ (right panel)
as a function of the partonic momentum fraction $x$.
For each PDF, the bands  provide an estimate of the associated uncertainty.
The dependence of the PDFs with the scale $Q$ is determined by the perturbative  DGLAP
evolution equations.
        \label{fig:dglap} }
  \end{center}
\end{figure}

While the discussion so far has been kept somewhat schematic, it 
provides an intuitive picture of
the inner workings of the global PDF analysis paradigm.
By combining a wide range
of experimental measurements that involve proton targets with state-of-the-art
theory calculations, it becomes possible to  determine the quark and gluon
substructure of the proton.
Fig.~\ref{fig:dglap}  presents
the results of a recent determination of the proton structure,
the NNPDF3.0 global analysis~\cite{Ball:2014uwa}.
This comparisons shows the up and down quark valence PDFs,
defined as $f_{u_V}=f_u - f_{\bar{u}}$ and $f_{d_V}=f_d - f_{\bar{d}}$,
the sea quark PDFs $f_{\bar{u}}$, $f_{\bar{d}}$,
$f_{s}$, $f_{c}$, $f_{b}$, as well as the gluon PDF $f_g$ (divided by 10).
The PDFs are displayed
both at $Q^2=10$ GeV$^2$ (left)
and at $Q^2=10^4$ GeV$^2$ (right panel)
as a function of the partonic momentum fraction $x$.
Recall that the dependence of the PDFs with the scale $Q$ is
entirely fixed by the perturbative  DGLAP
evolution equations.
For each of the PDFs shown in
Fig.~\ref{fig:dglap},
the size of the bands provides an estimate of the associated uncertainty.

%%%%%%%%%%%%%%%%%%%%%%%%%%%%%%%%%%%%%%%%%%%%%%%%%%%%%%%%%%%%%%%%%%%
%%%%%%%%%%%%%%%%%%%%%%%%%%%%%%%%%%%%%%%%%%%%%%%%%%%%%%%%%%%%%%%%%%%%%%

From Fig.~\ref{fig:dglap} one can observe that the overall
shape of the quark valence distributions, $f_{u_V}$
and $f_{d_V}$, is consistent with the condition that the total number of valence quarks in the proton
is two for the up and one for the down quarks.
This is a direct consequence of the valence sum rules of Eq.~(\ref{eq:valenceSumRules}).
Furthermore, one sees that at large values of the momentum fraction $x$ the valence quark
PDFs are the largest, while at smaller $x$ the proton is dominated by its sea quark
and gluon components.
Indeed, there is steep rise at small-$x$ in the number of gluons
and sea quarks, which quickly dominate over the valence distributions,
in particular as $Q$ is increased.
As one moves from $Q^2=10$ GeV$^2$ to $Q^2=10^4$ GeV$^2$  as determined by
the DGLAP evolution equations,
one finds that the valence quarks $f_{u_V}$ and $f_{d_V}$ are relatively
stable while the gluons and sea quarks rise strongly.
This behaviour implies that the higher the energies at which
one probes the
internal structure of the proton, the larger
its gluonic component will be.

The key property that allows the exploitation of the information
contained by the PDFs in different experiments is their
{\it universality}.
Thanks to the factorisation
theorems of the strong interaction, it is possible to determine the PDFs from a given type
of processes, such as lepton-proton scattering, and then use the
same PDFs to compute
predictions for different types of processes,
such as proton-proton collisions.
Fig.~\ref{fig:factorisation} illustrates one important application
of this PDF universality: by virtue
   of the QCD factorisation theorems, one can extract the proton PDFs
   from deep-inelastic
   scattering measurements (left) and then use the same PDFs
   to evaluate the production cross-section for
   electroweak gauge bosons in proton-proton collisions (right),
   the so-called Drell-Yan process.
   It is important to emphasize
   that this universality property is highly non-trivial:
   beyond the leading approximation, PDFs need to reabsorb
   into their definition higher-order effects that arise from
   the radiation of extra quarks and gluons.
Fortunately, it can be shown that such additional radiation effects are process-independent,
and thus do not compromise the universality of the PDFs.
      
\begin{figure}[h]
  \begin{center}
    \includegraphics[width=1.0\textwidth]{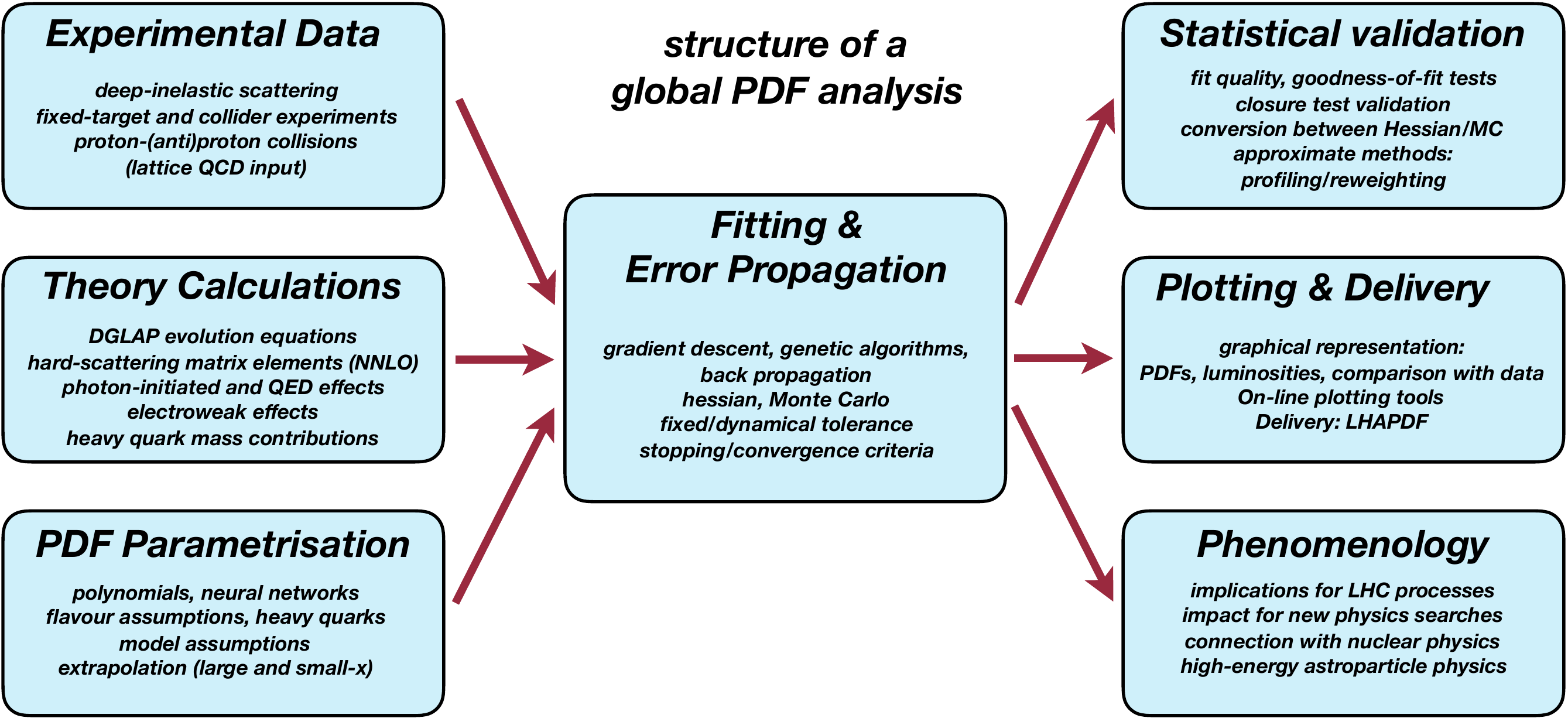}
    \caption{\small Schematic representation of the general
      structure of a global
      PDF analysis.
      It is based on three main groups
      of inputs: experimental data, theoretical calculations,
      and the PDF parametrisation and flavour assumptions.
      These inputs are processed and the most likely values of the PDFs
      and their associated uncertainties are returned.
      Subsequently, the results of this global PDF analysis are
      statistical validated,
      made publicly available, and their phenomenological
      implications for the LHC
      and other experiments studied.
        \label{fig:QCDanalysis} }
  \end{center}
\end{figure}

The technical complexities that
underlie a global PDF analysis are
schematically summarised in  Fig.~\ref{fig:QCDanalysis}.
It comprises three main types
of inputs: experimental measurements, theoretical calculations,
and the methodological assumptions such as those related
to the PDF parametrisation and the quark flavour decomposition.
This theory input involves higher-order perturbative calculations in both
the strong and electroweak interactions for the DGLAP evolution kernels
and for the hard-scattering matrix elements.
In addition, these higher-order calculations need to be interpolated
in the form of fast-access grids~\cite{Carli:2010rw,Wobisch:2011ij,amcfast}
in order to satisfy the requirements
of the CPU-time intensive PDF fits.

These inputs are processed through a fitting program that
returns the most likely values of the PDFs
and their associated uncertainties.
Several methods have been developed to estimate
PDF uncertainties, with the Hessian~\cite{Pumplin:2001ct}
and the Monte Carlo~\cite{DelDebbio:2007ee} approaches being the most popular.
In addition, approximate techniques have been constructed to emulate within certain
approximations the results of a full PDF fit in a much faster way, in particular
the Bayesian reweighting of Monte Carlo sets~\cite{Ball:2011gg}
and the profiling of Hessian sets~\cite{Paukkunen:2014zia}.
Subsequently, the results of this global PDF analysis are
statistically validated,
made publicly available, and their phenomenological
implications for the LHC
and other experiments studied.
It is beyond the scope of this Article to describe each
of the components listed in Fig.~\ref{fig:QCDanalysis}, for further
detail the reader could consult~\cite{Gao:2017yyd} and references therein.

%%%%%%%%%%%%%%%%%%%%%%%%%%%%%%%%%%%%%%%%%%%%%%%%%%%%%%%%%%%%%%%%%%%
%%%%%%%%%%%%%%%%%%%%%%%%%%%%%%%%%%%%%%%%%%%%%%%%%%%%%%%%%%%%%%%%%%%%%%

Global fits of parton distributions, such as the one displayed
in Fig.~\ref{fig:dglap}, have played
a central role in the interpretation of experimental measurements
in lepton-hadron and hadron-hadron collisions in the last three decades.
As highlighted by Fig.~\ref{fig:factorisation}, any theoretical predictions
for processes at colliders that involve protons in the initial state,
such as HERA and the LHC, necessarily require PDFs as input.
Parton distributions in particular were one of the theoretical inputs
that contributed to the
discovery of the Higgs boson in 2012 by the ATLAS and CMS
experiments~\cite{Aad:2012tfa,Chatrchyan:2012xdj}, recognised with the Nobel Prize in Physics in 2013,
and which heralded a new era for elementary particle physics.
The Higgs boson can be rightly qualified as the most spectacular particle ever encountered.
First of all, it is the  only known particle which couples to anything that has mass.
Second, it transmits a new type of fundamental force, which is completely
different to all other interactions such as electromagnetism.
Thirdly, it is exquisitely  sensitive to quantum effects
taking place at the tiniest of the distances.
Nowadays, improving our understanding of the quark and gluon structure
of the proton makes possible
scrutinizing the Higgs sector of the Standard Model in greater detail, and indeed several
recent developments in the field of PDF fits aim to strengthen the robustness of
theoretical predictions for Higgs boson production at the LHC~\cite{deFlorian:2016spz}.

So far the discussion in this Article
has been restricted to the partonic structure of free protons.
A closely related field of research focuses
on the study of the partonic structure of nucleons (protons and neutrons)
which are bound within heavy nuclei.
The quark and gluon structure of bound nucleons
is parametrised by the so-called nuclear parton distribution functions (nPDFs).
It was discovered by the EMC experiment in the 1980s~\cite{Aubert:1983xm},  to great surprise,
that the parton distributions of bound nucleons were significantly different
from those of their free-nucleon counterparts.
Such behaviour was quite unexpected, since nuclear binding effects are characterised
by MeV-scale momentum transfers, which should be negligible 
when compared to the typical momentum transfers ($Q\gsim 1$ GeV) involved
in  deep-inelastic scattering.

\begin{figure}[h]
  \begin{center}
    \includegraphics[width=0.35\textwidth,angle=-90]{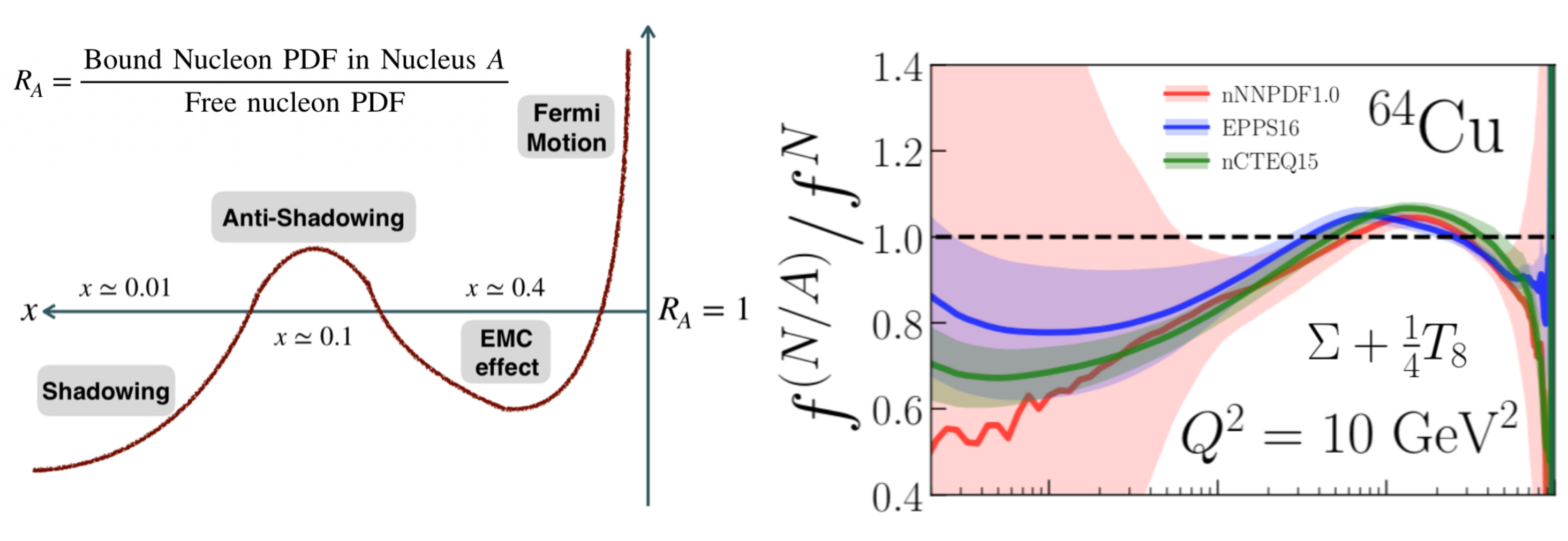}
    \caption{\small Left: schematic representation of nuclear effects that modify the PDFs
      of nucleons bound within heavy nuclei (nPDFs) as compared to their free nucleon
      counterparts.
      Right: comparison of the nuclear modifications for the $\Sigma+T_8/4$ quark
      combination in copper ($A=64$) between three recent fits of nuclear PDFs at $Q^2=10$
      GeV$^2$, where the bands represent the associated nPDF uncertainties.
        \label{fig:npdfs} }
  \end{center}
\end{figure}

In addition to this suppression of the nPDFs at $x\simeq 0.4$ (the EMC effect), other
nuclear modifications that have been reported include an enhancement at large-$x$
(Fermi motion) and a further suppression in the small-$x$ region
known as nuclear shadowing.
The left panel of Fig.~\ref{fig:npdfs} displays
a schematic representation of how different nuclear effects modify the PDFs
of nucleons bound within heavy nuclei as compared to those of free nucleons
as a function of the partonic momentum fraction $x$.
Note that in such comparisons
the limit $R_A=1$ corresponds to the absence of nuclear corrections.

%%%%%%%%%%%%%%%%%%%%%%%%%%%%%%%%%%%%%%%%%%%%%%%%%%%%%%%%%%%%%%%%%%%
%%%%%%%%%%%%%%%%%%%%%%%%%%%%%%%%%%%%%%%%%%%%%%%%%%%%%%%%%%%%%%%%%%%%%%

From the conceptual and methodological points of view, global fits of nuclear
PDFs proceed in a similar way as those of the free nucleon PDFs and are summarised in
Fig.~\ref{fig:QCDanalysis}.
This said, there are two  important differences compared to the free proton case.
First of all, the available experimental constraints are greatly reduced in
the nuclear case, with a restricted kinematical coverage both in $x$ and in $Q^2$.
Secondly, one needs to extend the fitting methodology to also account  for the dependence on
the nuclear mass number $A$.
This implies that nPDFs will depend on three variables, $f_i^{(A)}(x,Q^2,A)$,
two of which ($x$ and $A$) are determined by non-perturbative dynamics and thus
need to be parameterised and extracted from data.
Several groups have presented fits of nPDFs with varying input datasets, theory
calculations, and methodological assumptions.
The right plot of Fig.~\ref{fig:npdfs} displays
a comparison of the nuclear modifications for the $\Sigma+T_8/4$ quark
combination\footnote{Where one has defined $f_\Sigma=\sum_i \lp f_{q_i}+f_{\bar{q}_i}\rp$
  and $f_{T_8}= f_{u_i}+f_{\bar{u}_i}+f_{d_i}+f_{\bar{d}_i}-2(f_{s_i}+f_{\bar{s}_i}) $. }
in copper ($A=64$) at $Q=10$ GeV$^2$ between the nNNPDF1.0~\cite{AbdulKhalek:2019mzd},
nCTEQ15~\cite{Kovarik:2015cma}, and EPPS16~\cite{Eskola:2016oht}
nPDF sets, where the bands indicate the associated uncertainties.
While there is reasonable agreement in the central values, there are marked differences
in the uncertainty estimates, in particular in the small and large-$x$ regions.

In addition to shedding light on the inner workings of the strong force
in the nuclear environment,
the accurate determination of nuclear PDFs represents an important
ingredient for the interpretation of the results
of the heavy ion programs of RHIC and the LHC, where one collides
heavy nuclei such as lead ($A=208$).
In such heavy ion collisions, it becomes possible to study the properties of the
Quark Gluon Plasma (QGP), the hot and dense medium created  in the early
Universe and that now can be replicated in the laboratory.
Nuclear PDFs enter the initial state of heavy ion collisions
whenever
hard probes such as jets, weak bosons, or heavy quarks are produced~\cite{Abreu:2007kv}.
Therefore, improving our understanding of the nPDFs is important in order 
discriminate between
the cold from the hot nuclear matter effects in those complex events, involving hundreds
or even thousands of produced particles.

Following this general introduction to the topic of the partonic
structure of nucleons and nuclei,
now  an overview of the current state
of (n)PDF determinations is presented, emphasizing recent breakthroughs and important results
in the field.

%Part Two (3500 – 4500 words)
% Present the current state of the science, discipline or areas of study that your Article focuses on, including strengths and weaknesses. Include observational, theoretical and experimental techniques used.
% Refer to work in as many other countries as is sensible.
% You may add material from your own research in moderation.

\section{The partonic structure of hadrons in the LHC era}

\paragraph{Global PDF fits.}
Several collaborations provide regular updates of their global proton PDF analyses,
including ABMP~\cite{Alekhin:2017kpj}, CT~\cite{Hou:2019qau}, MMHT~\cite{Harland-Lang:2014zoa}, and
NNPDF~\cite{Ball:2017nwa}.
The differences among them stem from several of the aspects that define a PDF fit
(see Fig.~\ref{fig:QCDanalysis}).
To begin with, the input datasets are not the same, for example ABMP does not include
jet production measurements, and not all groups include the same neutrino DIS
experiments.
Secondly, theoretical calculations can differ, for instance, part of the disagreement between ABMP
and other groups can be traced back to the treatment of the heavy quark corrections
in DIS structure functions.
Furthermore, the treatment of the Standard Model inputs in the PDF analysis
is not homogeneous: while CT and NNPDF take external
parameters such as the strong coupling constant $\alpha_s(m_Z)$ and
the charm and bottom quark masses $m_c$ and $m_b$ from their Particle
Data Group averages~\cite{Olive:2016xmw}, the ABMP and MMHT groups
determine these parameters instead
simultaneously with the PDFs.

\begin{figure}[h]
  \begin{center}
    \includegraphics[width=0.99\textwidth]{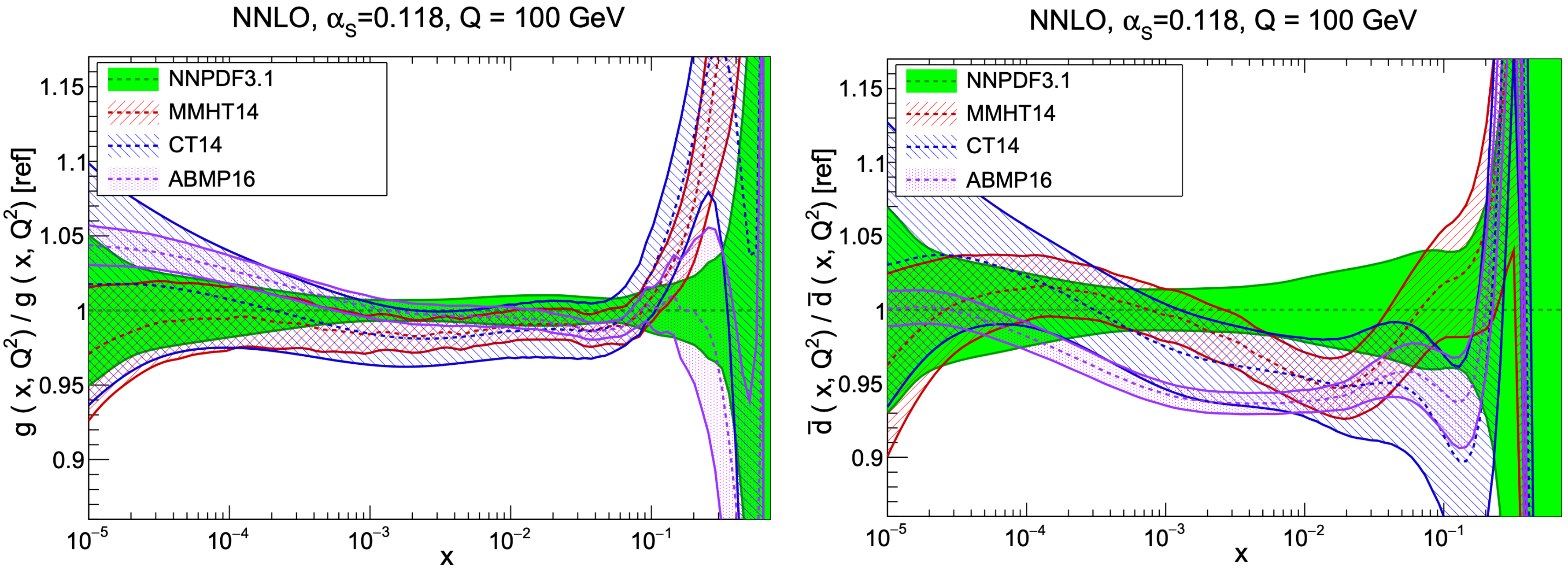}
    \caption{\small Comparison between the gluon (left) and down antiquark
      (right plot) PDFs at $Q=100$ GeV between the ABMP16, MMHT14, CT14, and NNPDF3.1
      fits, normalised to the central value of the latter.
      For each fit, the bands represent the 68\% confidence level uncertainties.
        \label{fig:globalfits} }
  \end{center}
\end{figure}

The third, and perhaps most important, source
of differences between the PDF fitting groups are related to
methodological assumptions such as the way
to parametrise PDFs and to estimate the associated uncertainties.
Here there are two main groups: ABMP, CT, and MMHT are based on the Hessian method
(with or without tolerances) and a polynomial parametrisation, while NNPDF adopts the Monte Carlo
method with artificial neural networks as universal unbiased interpolants.
Fig.~\ref{fig:globalfits} compares the gluon and the down antiquark
PDFs at $Q=100$ GeV between the ABMP16, MMHT14, CT14, and NNPDF3.1
fits, normalised to the central value of the latter.
For each fit, the bands represent the 68\% confidence level uncertainties.
Similar comparison plots (also for related
quantities such as PDF luminosities) can be straightforwardly produced by means
of the APFEL-Web online PDF plotter~\cite{Carrazza:2014gfa}.

%%%%%%%%%%%%%%%%%%%%%%%%%%%%%%%%%%%%%%%%%%%%%%%%%%%%%%%%%%%%%%%%%%%
%%%%%%%%%%%%%%%%%%%%%%%%%%%%%%%%%%%%%%%%%%%%%%%%%%%%%%%%%%%%%%%%%%%%%%

\paragraph{Impact of LHC data.}
Traditionally, hadron colliders were considered the domain of discovery physics, while the cleaner
lepton colliders had the monopoly of precision physics.
However, the outstanding performance of the LHC and its experiments, complemented
by the recent progress
in higher order QCD and electroweak calculations, have demonstrated that despite this common lore
the LHC is able to pursue a vigorous and exciting precision physics program.
In the context of studies of the partonic content of nucleons and nuclei, this precision program
aims to exploit the information contained in LHC measurements to pin down the quark
and gluon PDFs and to reduce their uncertainties.
In turn, this should
lead to improved theory predictions
for many important processes at the LHC and elsewhere, ranging 
from Higgs production to Dark Matter searches.

Until around 2012, all PDF fits were restricted to deep-inelastic scattering
measurements (from both fixed-target experiments and from the HERA collider), fixed-target
Drell-Yan cross-sections, and some jet production measurements
from the Tevatron.
Fortunately the situation has improved dramatically in the recent years.
A large number of  LHC processes are now staple components of
global PDF analyses: differential Drell-Yan measurements,
the transverse momentum of $W$ and $Z$ bosons, top quark pair and single
top production, and heavy meson production, among several others.
In most cases, state-of-the-art NNLO QCD calculations are available,
allowing fully consistent NNLO PDF fits based
on a the widest possible range of fixed-target and collider measurements to be carried out.

%%%%%%%%%%%%%%%%%%%%%%%%%%%%%%%%%%%%%%%%%%%%%%%%%%%%%%%%%%%%%%%%%%
\begin{figure}[h]
  \begin{center}
    \includegraphics[width=0.99\textwidth]{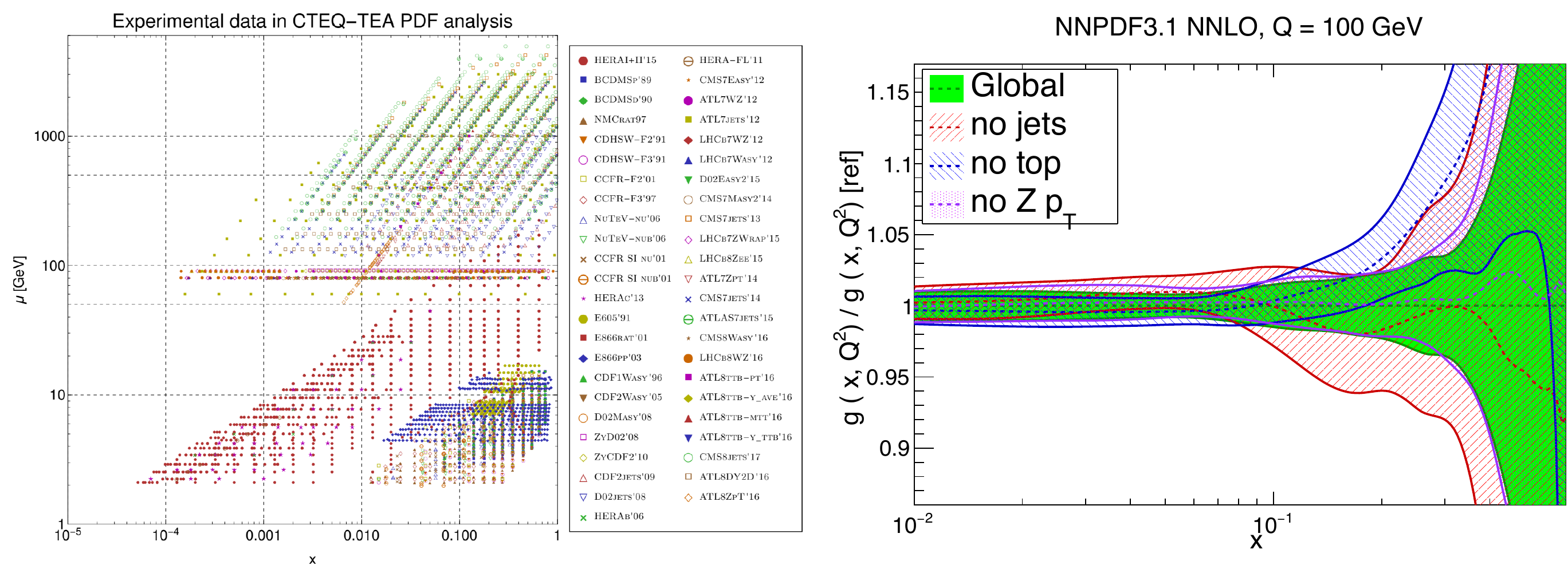}
    \caption{\small Left: the kinematic coverage in the $(x,Q)$ plane
      of the experimental data included in a recent CT analysis.
      Right: comparison of the large-$x$ gluon from the global NNPDF3.1 fit
      with the corresponding results in fits without the jet, top quark, or
      the $Z$ $p_T$ measurements
      included.
        \label{fig:kinlhc} }
  \end{center}
\end{figure}
%%%%%%%%%%%%%%%%%%%%%%%%%%%%%%%%%%%%%%%%%%%%%%%%%%%%%%%%%%%%%%%%%%%%%%

In order to illustrate the breadth of collider measurements included
in a typical modern PDF fit,
the left panel of Fig.~\ref{fig:kinlhc} displays
the kinematic coverage in the $(x,Q)$ plane
of the experimental data included in the recent CT analysis of~\cite{Hobbs:2019gob}.
One can observe that this coverage extends down to $x\simeq 5\times 10^{-5}$
and up to several TeV in $Q$, and furthermore that in most cases several
processes constrain the PDFs in the same region of the $(x,Q)$ plane.
Around half of the datasets
in this fit correspond to LHC measurements.
One important advantage of this rich dataset is the redundancy is provides,
meaning that a given PDF will typically be constrained by several datasets,
resulting in a significantly more robust PDF extraction.
This important property is illustrated in the case of the large-$x$ gluon
in the right panel of Fig.~\ref{fig:kinlhc},
which displays the results from the global NNPDF3.1 NNLO fit
compared with the corresponding results in fits without any jet, top quark, or
$Z$ $p_T$ measurements included.
One observes that in all cases the resultant gluon PDFs are consistent
within uncertainties, highlighting the complementarity of the various gluon-sensitive
measurements provided by the LHC.

\paragraph{PDFs with theory uncertainties.}
An aspect of the global PDF fitting machinery that has attracted a lot of attention
is the estimate and propagation of the associated uncertainties.
In most PDF fits, the uncertainty estimate
accounts only for the propagation of the uncertainties corresponding
to the input experimental measurements, as well as for methodological
components associated  with {\it e.g.} the choice of functional form.
PDF analyses however do not account in general for
the theory uncertainties associated with the truncation of
the perturbative expansions of the input QCD calculations.
While neglecting these theory errors might have been justified in the past,
the rapid pace of improvements in PDF fits, leading to ever smaller
uncertainties, suggested that such assumption should be revisited.

In this context, an important recent development in global PDF analyses
has been the formulation
of frameworks to systematically include theoretical
uncertainties,
in particular those associated to the missing higher orders (MHOs) in
the perturbative QCD calculations used as input to the fit.
One example is the method proposed in~\cite{AbdulKhalek:2019bux,AbdulKhalek:2019ihb},
whose basic idea is constructing a combined covariance
matrix that includes both experimental and theoretical components,
with the latter estimated using the scale-variation method and validated
whenever possible with existing higher order calculations.
It is important to mention that other approaches to estimate
the impact of MHOUs in PDF fits have been presented elsewhere,
for example~\cite{Harland-Lang:2018bxd}  proposes a consistent use of scale variations in PDF fits
and the corresponding physical predictions.
More research is needed in this important topic, and in particular perhaps
the scale variation approach turns out not to be the optimal strategy to estimate
the impact of MHOUs in PDF fits.

\begin{figure}[h]
  \begin{center}
    \includegraphics[width=0.99\textwidth]{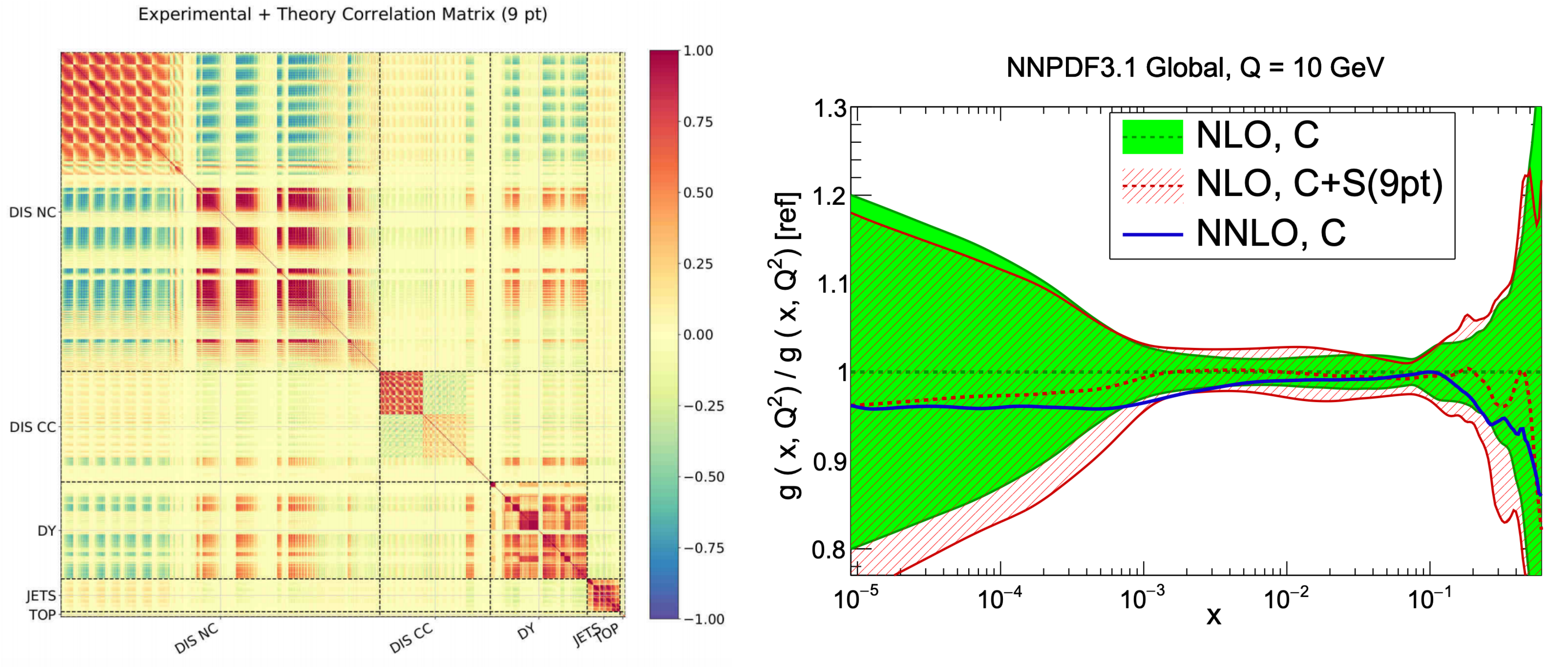}
    \caption{\small Left: the combined experimental and theoretical
      correlation matrix for the dataset used in the NNPDF3.1 analysis.
      Note how theory-induced correlations now connect different
      measurements, such as DIS and Drell-Yan, which are
      experimentally uncorrelated.
      Right: the results of the NNPDF3.1 NLO fit with theory
      uncertainties, compared with a baseline with only experimental
      errors in the covariance matrix
      and with the central value of the corresponding NNLO fit.
        \label{fig:thcovmat} }
  \end{center}
\end{figure}

Within the theory covariance matrix approach developed by the NNPDF group, beyond
an overall increase of the total uncertainties, is that
theory-induced correlations now connect different
measurements, such as DIS and Drell-Yan or top quark
production, which are
experimentally uncorrelated.
This feature is illustrated by the  left panel of Fig.~\ref{fig:thcovmat}, which displays
the combined experimental and theoretical
 correlation matrix for the dataset used in the NNPDF3.1 NLO analysis.
 There one sees for example that theory uncertainties introduce a positive correlation
 between DIS and Drell-Yan in some kinematic regions, and a negative one in others.

 %%%%%%%%%%%%%%%%%%%%%%%%%%%%%%%%%%%%%%%%%%%%%%%%%%%%%%%%%%%%%%%%%%%
%%%%%%%%%%%%%%%%%%%%%%%%%%%%%%%%%%%%%%%%%%%%%%%%%%%%%%%%%%%%%%%%%%%%%%

 This formalism has been used to carry out an updated version
of the NNPDF3.1 NLO global analysis now accounting for the MHOUs.
The right panel of Fig.~\ref{fig:thcovmat}
displays  the results of the NNPDF3.1 NLO fit with theory
uncertainties (red hatched band) for the gluon PDF at $Q=10$ GeV,
compared with a baseline with only experimental
errors in the covariance matrix (green solid band)
and with the central value of the corresponding NNLO fit.
There are two main consequences of including the theory
uncertainties in the fit covariance matrix.
First of all, a moderate increase of the PDF uncertainties, showing that theory
errors cannot be neglected when estimating the total PDF uncertainties, at least at NLO.
Second, a shift in the central values due to the rebalancing effect
of theory errors and their correlations, which  in general
are found to shift
the fit results towards the corresponding NNLO result.
The latter is a desirable feature of any method that estimates theory
uncertainties, whose ultimate goal is to accurately predict the results of the next
perturbative order.

\begin{figure}[h]
  \begin{center}
    \includegraphics[width=0.99\textwidth]{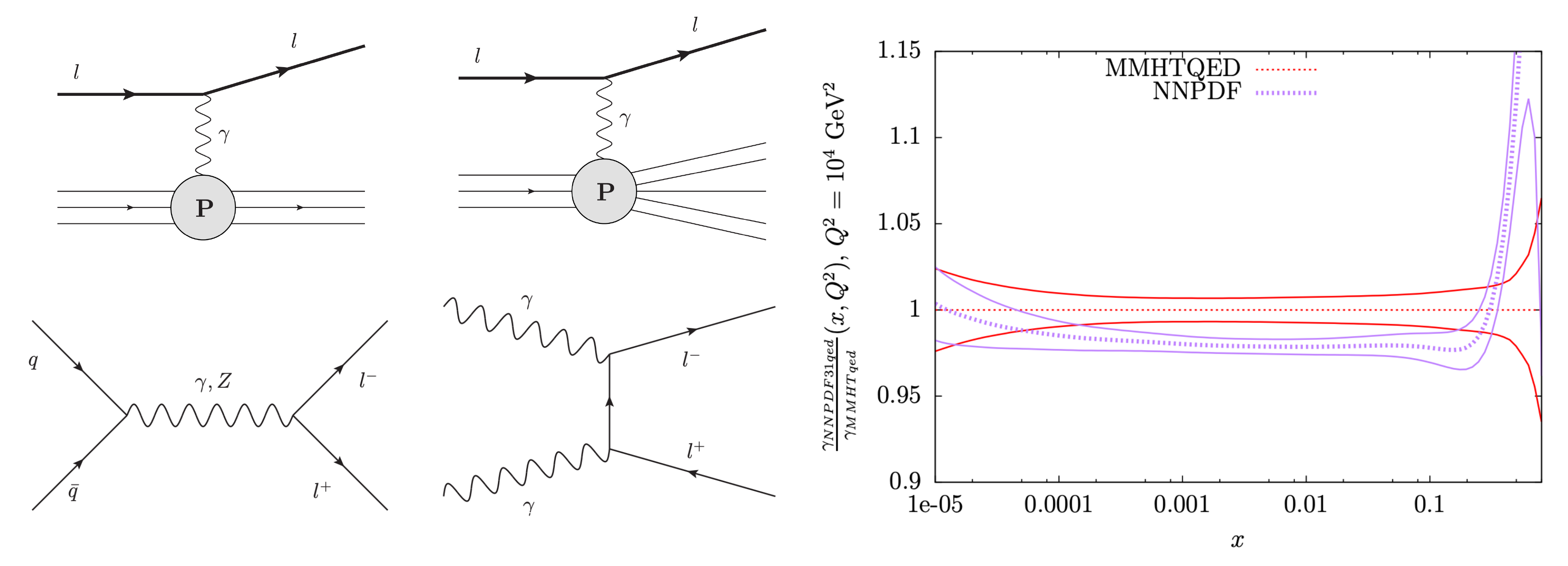}
    \caption{\small Upper left: the $F_{2,L}$ deep-inelastic structure functions have associated
      photon-induced contributions from elastic (where the proton remains intact)
      and inelastic processes (where the proton breaks down),
      from~\cite{Harland-Lang:2019pla}.
      Bottom left: photon-induced processes contribute to the total cross-sections
      of many important hadron collider processes, in this case Drell-Yan lepton-pair
      production.
      Right: comparison between the photon PDFs $f_{\gamma}(x,Q=100~{\rm GeV})$
      between the NNPDF and MMHT QED analysis, normalised to the central value
      of the latter.
        \label{fig:qed} }
  \end{center}
\end{figure}

\paragraph{The photon PDF.}
A perhaps unexpected observation whose implications
were only fully appreciated recently is related to
the fact that the partonic content of the proton
should be composed not only of quarks and gluons, but also of photons.
Indeed, it can be demonstrated that protons radiate quasi-real photons
that induce photon-initiated scattering reactions in lepton-proton
and in proton-proton collisions, and that these reactions can be
described as arising from a photon PDF $f_{\gamma}(x,Q^2)$.
This photon PDF
 receives two types of contributions,
as schematically indicated in the upper left panel of Fig.~\ref{fig:qed}:
from elastic processes, where the proton is left intact, and from inelastic
processes, where the proton breaks up after the scattering.

%%%%%%%%%%%%%%%%%%%%%%%%%%%%%%%%%%%%%%%%%%%%%%%%%%%%%%%%%%%%%%%%%%%
%%%%%%%%%%%%%%%%%%%%%%%%%%%%%%%%%%%%%%%%%%%%%%%%%%%%%%%%%%%%%%%%%%%%%%

The presence of a photon component in the proton has two main
phenomenological implications.
First of all, the photon mixes with the quark and gluon PDFs via QED effects
in the DGLAP evolution, and thus modifies the results
of the latter as compared to the QCD-only case.
In particular, the photon subtracts from the quarks
and gluons a small amount of the total momentum of the proton,
which is currently
estimated to amount to a few permille.
Second, photon-initiated contributions result in the opening
of new channels for important hard-scattering processes.
An important example of this is shown in 
the bottom left panel of Fig.~\ref{fig:qed}: photon-initiated processes
contribute to the total Drell-Yan cross-section (lepton pair production) on the same footing
as the quark-antiquark annihilation reactions.

Until recently, the photon PDF was determined either from 
model assumptions~\cite{Martin:2004dh,Schmidt:2015zda} or
freely parametrised and constrained by experimental data~\cite{Ball:2013hta}.
None of these options were satisfactory: the former due to the bias associated
with the choice of model, the latter since the results were affected by 
large  uncertainties due to limited available constraints.
A major breakthrough was then the demonstration that
the photon content of the proton does not need neither to be modeled nor
to be fitted from data, but rather it can be 
 expressed in terms of the well-measured
inclusive deep-inelastic scattering structure functions
$F_2$ and $F_L$.
This result was originally formulated by means the equivalent photon
approximation~\cite{Martin:2014nqa,Harland-Lang:2016kog},
and subsequently placed on a more rigorous footing by the LUXqed formalism
developed in~\cite{Manohar:2016nzj,Manohar:2017eqh}.

Since these pioneering analyses were presented, other 
global PDF fitting groups have provided QED variants of their PDF sets
that include both QED corrections in the DGLAP evolution with
a photon PDF determined by means of the LUXqed calculation
or variations thereof~\cite{Bertone:2017bme,Harland-Lang:2019pla}.
Given the relatively tight constraints imposed by the LUXqed framework, the resulting
photon PDFs turn out to be quite similar.
To illustrate this, the right panel of  Fig.~\ref{fig:qed} displays
a comparison of the photon PDFs $f_{\gamma}(x,Q)$
at $Q=100$ GeV
between the NNPDF and MMHT QED analyses, normalised to the central value
of the latter.
The two photon PDFs agree at the few percent level in most of the relevant
$x$ range, with the exception of the large-$x$ regions where differences
can be as large as 15\%.

\paragraph{Implications for astroparticle physics.}
Another topic that has received considerable attention in the recent years has
been the interplay between proton structure studies
in nuclear and particle physics with those in astroparticle physics.
This interest was motivated by the realisation that improving our
understanding of the partonic
content of nucleons and nuclei was an important ingredient
for those theoretical predictions
relevant for the interpretation of high-energy astrophysics experiments.
This connection is particularly important
for neutrino telescopes, such as IceCube and KM3NET, which instrument
large volumes of ice and water as effective detectors of energetic neutrinos,
as well as for cosmic ray detectors such as Auger, which aim to detect the
most energetic particles in the Universe.

\begin{figure}[h]
  \begin{center}
    \includegraphics[width=0.99\textwidth]{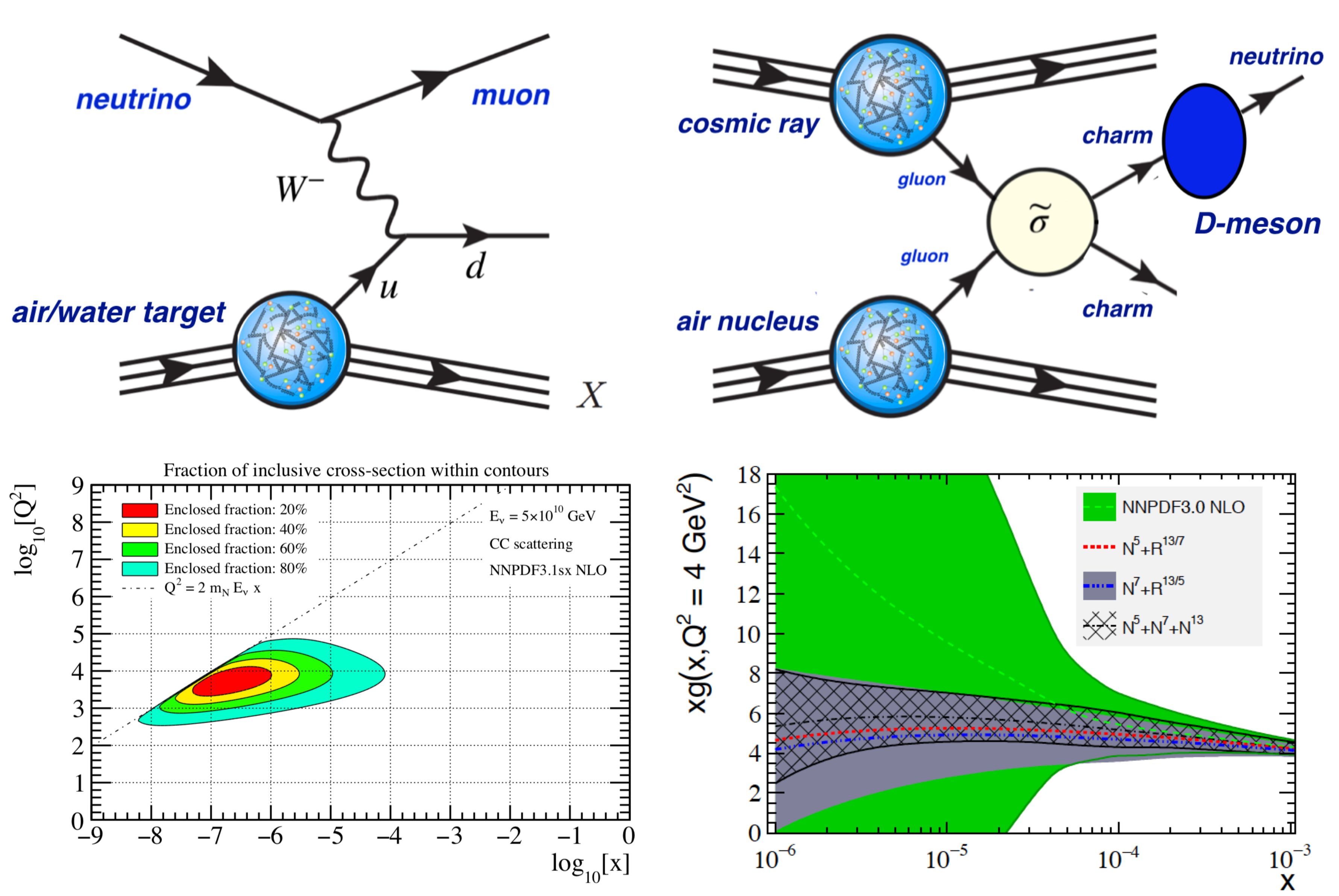}
    \caption{\small Upper plots: theoretical QCD calculations
      sensitive to the proton structure are required
      in high-energy neutrino astronomy both to predict signal event rates in
      neutrino detection (left panel) and the prompt
      neutrinos flux arising from cosmic ray collisions (right panel).
      Bottom left: the kinematic coverage in the $(x,Q^2)$ plane
      for charged-charged neutrino-nucleus scattering at
      an energy of $E_{\nu}=5\times 10^{10}$ GeV.
      Bottom right: the reduction in the PDF
      uncertainties of the small-$x$ gluon at $Q^2=4$ GeV$^2$
      once the LHCb charm production
      measurements are included in the NNPDF3.0 fit.
      \label{fig:neutrinos} }
  \end{center}
\end{figure}

In the specific case of high-energy neutrino astronomy, QCD calculations are
required for two different aspects of the data interpretation.
First of all, to predict the expected event rates of neutrino-nucleus interactions
at high energies
(upper left panel of Fig.~\ref{fig:neutrinos}) via the neutral-current
and charged current deep-inelastic
scattering processes.
Knowledge of these cross-sections is required
both to predict the number of neutrinos that will
interact with the nuclei in the ice or water targets, as well as the attenuation
of the incoming neutrino flux as they traverse through the Earth~\cite{Vincent:2017svp}.
As indicated in the bottom left panel of Fig.~\ref{fig:neutrinos},
for a neutrino with an energy of $E_{\nu}=5\times 10^{10}$ GeV the charged-current
DIS process probes the proton PDFs down to $x\simeq 10^{-8}$~\cite{CooperSarkar:2011pa},
a region far beyond present experimental constraints.
The interpretation of the measurements of neutrino telescopes
could thus be hindered by theoretical uncertainties in QCD and the partonic
content of protons.

Secondly, QCD processes are also relevant for the prediction of the dominant
background for neutrino astronomy, the so called prompt
neutrino flux.
In this process, represented in
the upper right panel of Fig.~\ref{fig:neutrinos},
an energetic cosmic ray, typically a proton, collides
with an air nucleus and produces a charmed meson, that prompt decays
into neutrinos.
The flux of such neutrinos becomes at high energies larger than those
from pion and kaon decays, since the short lifetime of $D$ mesons implies
that their energy is not attenuated before decay.
As in the neutrino cross-section case, the calculation of the prompt neutrino flux involves
knowledge of the proton structure for $Q\simeq m_c=1.5$ GeV down
to $x\simeq 10^{-6}$~\cite{Zenaiev:2015rfa}, where theoretical
uncertainties are large.

%%%%%%%%%%%%%%%%%%%%%%%%%%%%%%%%%%%%%%%%%%%%%%%%%%%%%%%%%%%%%%%%%%%
%%%%%%%%%%%%%%%%%%%%%%%%%%%%%%%%%%%%%%%%%%%%%%%%%%%%%%%%%%%%%%%%%%%%%%

The key ingredient allowing providing solid theoretical
predictions for both types of  processes is the exploitation
of collider measurements sensitive to the small-$x$ structure of the proton, such as
 the charm production data in the forward region from the LHCb experiment.
Forward $D$ meson production at LHCb, when Lorentz-boosted to the center-of-mass frame,
covers the same kinematic region as that of charm production
in energetic cosmic ray collisions, and therefore
makes it possible to pin down the gluon distribution
at small-$x$ beyond the reach of the HERA collider data and in the
kinematical region of relevance for neutrino astronomy.
The bottom right of  Fig.~\ref{fig:neutrinos} displays the
significant reduction in the PDF
uncertainties of the small-$x$ gluon once the LHCb charm production
measurements at $\sqrt{s}=5,7$ and 13 TeV are included in the NNPDF3.0 fit~\cite{Gauld:2016kpd}.
Thanks to this connection with LHC hard-scattering
processes, it has become possible to provide
robust state-of-the-art predictions for the signal~\cite{Bertone:2018dse} and
background~\cite{Bhattacharya:2016jce} processes at
neutrino telescopes with reduced theoretical uncertainties.
In the future, one can envisage a situation where measurements at high-energy
astroparticle physics experiments can be used to constrain the small-$x$
structure of the nucleon and the associated QCD dynamics in this regime.

\paragraph{PDFs and BSM searches.}
As more data is being collected and the precision and breadth of the LHC measurements improves,
the information on the partonic structure of
the proton that one can obtain from them will be consequently increased.
In particular, the large integrated luminosities that
will be accumulated in future runs of the LHC
imply that extended measurements in
the TeV region should make possible a more precise
determination of poorly known PDFs such as the large-$x$ sea quark and the gluon
distributions,
see also the left panel of Fig.~\ref{fig:kinlhc}.
A potential worry in using LHC data in this high-energy region
in PDF fits is that the fits could be biased  if deviations with respect
to the SM were present.
This concern is becoming more acute due to the
lack of new particles or interactions detected so far at the LHC:
this might suggest the presence of a band
gap between the electroweak scale and the scale of new physics $\Lambda$.
In turn, such a band gap could imply that Beyond the Standard Model (BSM) dynamics
could very well manifest itself at the LHC only via subtle deviations
in the tails of the measured distributions.

A powerful framework to interpret in a model-independent manner
the results of the LHC is provided by the Standard Model Effective Field Theory (SMEFT)~\cite{Brivio:2017vri}.
Within the mathematical language of the SMEFT, the effects of BSM dynamics at high energies
$\Lambda \gg v$ above the electroweak scale ($v=246$ GeV)
can be parametrised at lower energies, $E\ll \Lambda$, in terms of higher-dimensional
operators built up from Standard Model fields
and satisfying its symmetries such as gauge invariance.
Since several of the processes that are used at the LHC to constrain the SMEFT
degrees of freedom
are also used as inputs to the PDF determinations, it is important to assess their interplay
and to establish whether one could use the global PDF fits to disentangle
BSM effects from QCD dynamics, as originally proposed in~\cite{Berger:2010rj}.

\begin{figure}[h]
  \begin{center}
    \includegraphics[width=0.99\textwidth]{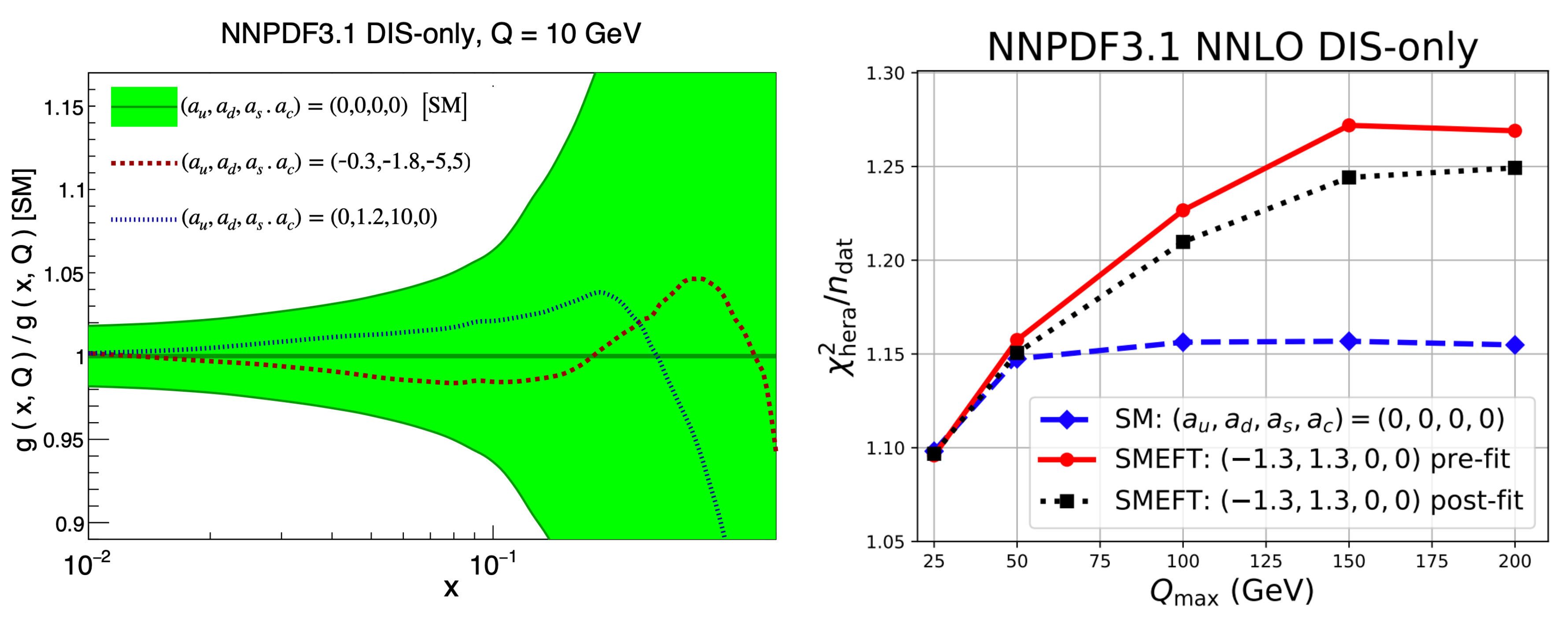}
    \caption{\small Left: the gluon PDF at large-$x$ for $Q=10$ GeV,
      comparing the results of fits based on SM calculations with those
      where the SM has been extended by specific combinations of SMEFT
      $d=6$ operators.
      Right: the different scaling of the $\chi^2$ with the energy of the process
      can be used to disentangle QCD effects from genuine BSM
      dynamics, in the case shown for a specific SMEFT benchmark point.
      \label{fig:smeft} }
  \end{center}
\end{figure}

A first study in this direction has been
recently presented in~\cite{Carrazza:2019sec},
where variants of the NNPDF3.1 DIS-only fit were carried out using theory
calculations where the SM had been extended by specific subsets of dimension-6
four-fermion SMEFT operators.
The left panel of Fig.~\ref{fig:smeft} shows
the results of the gluon PDF at large-$x$ for $Q=10$ GeV
in those fits.
For benchmark points in the parameter space not already excluded by other
experiments, one finds that the shift due to SMEFT corrections in the theory calculation
is at most half of the PDF uncertainty.
Crucially, one can exploit how the value of the fit-quality $\chi^2$ changes
with the energy of the process
to disentangle QCD effects from genuine BSM
dynamics, as show in the right panel of Fig.~\ref{fig:smeft}: the former
are smooth as the energy increases, since DGLAP evolution effects
are logarithmic in $Q^2$, while the latter are much
more marked since they scale as a power of $Q^2$.
While this study found that SMEFT-induced distortions
are sub-dominant with respect to PDF uncertainties, it was restricted
to DIS measurements and the picture could change significantly in the case
of LHC measurements, specially the high-statistics ones from future LHC runs.

%%%%%%%%%%%%%%%%%%%%%%%%%%%%%%%%%%%%%%%%%%%%%%%%%%%%%%%%%%%%%%%%%%%
%%%%%%%%%%%%%%%%%%%%%%%%%%%%%%%%%%%%%%%%%%%%%%%%%%%%%%%%%%%%%%%%%%%%%%

\paragraph{Constraints on nPDFs from LHC data.}
Concerning nuclear PDFs, perhaps
the most important recent development has been the availability of a wide variety of
hard probe measurements from proton-lead collisions at the LHC.
Similar to the proton PDF case, these measurements
provide new and valuable constraints on various nPDF combinations,
in particular for those that
can only be loosely constrained from DIS data such as the gluon and specific quark
flavour PDFs.
The information on the nuclear modifications of the gluon PDF
is particularly valuable, since these
are essentially unconstrained if only DIS data is
used in the nPDF fit.
Furthermore, these hard-scattering
LHC measurements offer novel opportunities to test the
validity of both the nPDF universality and
of the QCD factorisation properties in the nuclear environment.
Of specific interest is the small-$x$ regime, where eventual non-linear saturation
dynamics are expected to be enhanced as compared to the free-nucleon case.

\begin{figure}[h]
  \begin{center}
    \includegraphics[width=0.99\textwidth]{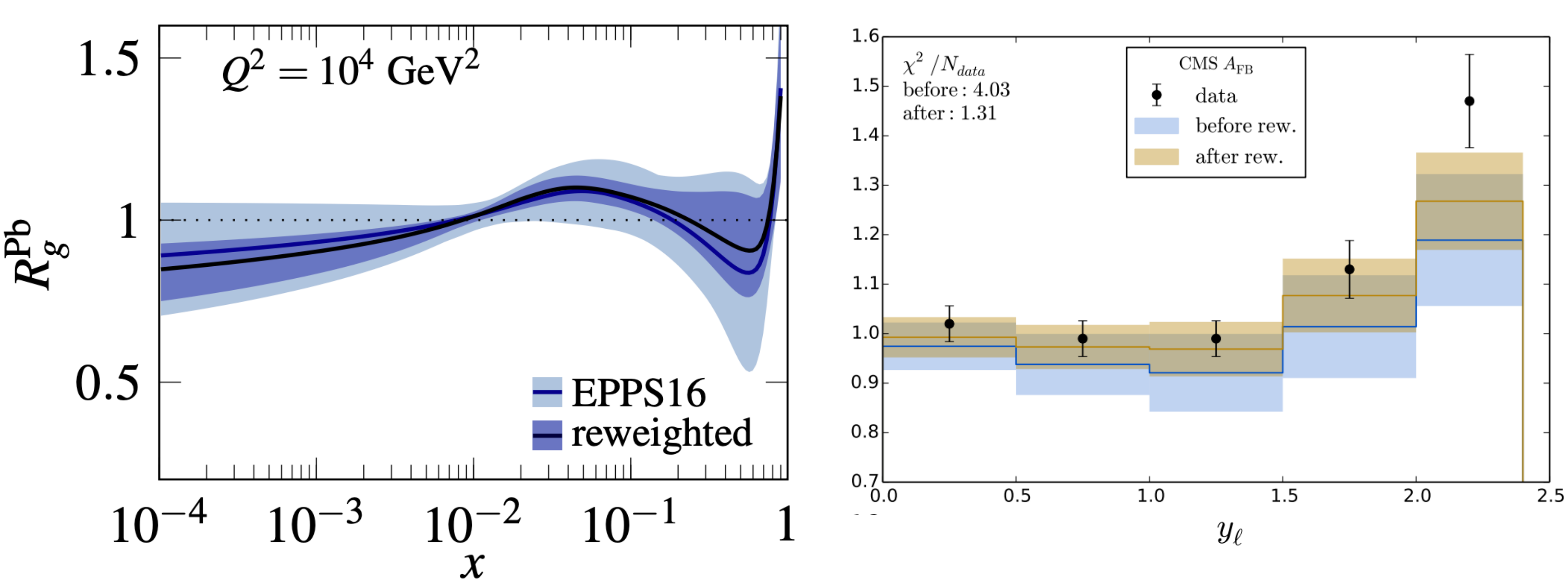}
    \caption{\small Left: the gluon nuclear ratio $R_g^{\rm Pb}$ 
      in lead at $Q=100$ GeV in the EPPS16 analysis before and after
      including the constraints from the CMS dijet measurements
      at $\sqrt{s}=5.02$ TeV via Hessian profiling.
      Right: the forward-backward asymmetry in $W$ production in proton-lead
      collisions from CMS, compared with the nCTEQ15 predictions before and
      after including this measurement in their fit.
        \label{fig:nuclearLHC} }
  \end{center}
\end{figure}

LHC measurements
from proton-lead collisions that have demonstrated their constraining power for
the determination of nPDFs
include dijet production~\cite{Eskola:2019dui} (sensitive to the gluon), $D$ meson production in the forward
region from LHCb~\cite{Kusina:2017gkz} (to pin down the small-$x$ gluon nPDF shadowing),
and electroweak gauge boson production (providing a handle on
the different quark flavour distributions).
Fig.~\ref{fig:nuclearLHC} illustrates the information on the nPDFs provided by hard
probes in proton-lead collisions from the LHC.
In the left panel the nuclear ratio $R_g^{\rm Pb}$ for gluons
in lead at $Q=100$ GeV is shown in the EPPS16 analysis before and after
including the constraints from the CMS dijet measurements
at $\sqrt{s}=5.02$ TeV via Hessian profiling.
One can observe how these data reduce the uncertainties of  $R_g^{\rm Pb}$
for a wide range of $x$ values, and in particular they appear to suggest
the presence of gluon shadowing at small-$x$.
Then the right panel of Fig.~\ref{fig:nuclearLHC}
 displays the  forward-backward asymmetry in $W$ production in proton-lead
collisions from CMS, compared with the nCTEQ15 predictions before and
after including this measurement in their fit.     
The agreement between data and theory is clearly improved
once the data on $A_{FB}$ is added to the analysis,
with $\chi^2/n_{\rm dat}$ decreasing from 4.03 to 1.31,
and also the uncertainties in the theory prediction
(which arise from the flavour decomposition of the quark nPDF)
are similarly reduced.

%%%%%%%%%%%%%%%%%%%%%%%%%%%%%%%%%%%%%%%%%%%%%%%%%%%%%%%%%%%%%%%%%%%
%%%%%%%%%%%%%%%%%%%%%%%%%%%%%%%%%%%%%%%%%%%%%%%%%%%%%%%%%%%%%%%%%%%%%%

% Conclusion (400 – 500 words)
% Draw together significant conclusions that assess the field, including strengths and weaknesses.
% Conclude with your judgment on what significant questions remain, are being pursued, or
%should be pursued.

\section{Summary and outlook}

As discussed throughout this Article,
the study of the partonic structure of nucleons and nuclei is an exciting and rich
field at the cross-roads between different aspects of particle, nuclear, and astroparticle
physics.
On the one hand, it allows us to shed light on the dynamics of the strong nuclear force, providing
crucial input on pressing questions such as the origin of the mass and the spin of protons,
the possible onset of new gluon-dominated regimes at small-$x$, the strange
and heavy quark contribution to the nucleon's wave function, or the dynamics
of quark and gluons for nucleons bound within heavy nuclei.
On the other hand, it provides essential input for the theoretical predictions
in processes as diverse as the production of Higgs bosons at the LHC,
the interactions of high-energy neutrinos at IceCube, and the collisions between
lead nuclei at the Relativistic Heavy Ion Collider.

In the recent years, the field has witnessed a number of important results
which highlight its vitality and productivity.
Just to mention a few of these:
the demonstration of the impact on the proton structure of precision LHC measurements;
the improved determinations of the strange, charm, and photon content of the
proton;
the first evidence for BFKL small-$x$ dynamics in HERA data~\cite{Ball:2017otu};
the calculation of $x$-space distributions with lattice QCD;
the formulation of new frameworks to estimate
and propagate theory uncertainties;
the establishment of the connection with neutrino telescopes
and cosmic ray physics;
and the analysis of the interplay between proton structure
and direct and indirect searches for New Physics at the high-energy frontier.

Many of these achievements in our
understanding of the partonic structure
of nucleons and nuclei have only become possible thanks to progress
from the methodological side,
from the development of new Machine Learning algorithms
to parameterize and train the PDFs~\cite{Carrazza:2019mzf} to strategies to combine and compress
different sets of parton distributions~\cite{Gao:2013bia,Carrazza:2015hva}
and new methods to quantify and represent
graphically the information provided by individual datasets~\cite{Wang:2018heo}.
Nevertheless, this list is necessarily incomplete due to space restrictions
of this Article, and the interested reader is encouraged
to turn to the more extended technical reviews mentioned in the introduction.

While great progress has been made in addressing long-standing questions
in our knowledge of the partonic structure of the proton,
important open questions still remain which should be addressed
in the coming years.
A major bottleneck in this respect will be to resolve some apparent incompatibilities
that affect current PDF interpretations of high-precision LHC measurements
for processes such as jet, top quark, and gauge boson production.
With statistical uncertainties at the few permille level in many cases,
confronting theory calculations with the experimental data requires dealing
with several hitherto ignored effects, such as the dependence
of the results on the experimental correlation model and reported tensions
between different processes and different distributions within the same process.
Addressing and overcoming these issues will be crucial to maximally exploit
the information contained in LHC data for precision PDF determinations.

To conclude this Article,  two possible future directions
for the field are highlighted.
The first one focuses on the exploitation of the constraints that will be provided
by PDF-sensitive measurements at future facilities, some of them already
approved, such as the High-Luminosity LHC, and some others still under
discussion, such as the Electron Ion Collider or the Large Hadron electron Collider.
The second direction points towards
the unification of different aspects of the global QCD analysis paradigm
into a unified framework that combines them in  a fully consistent way.

In order to further improve our understanding of the proton structure, several future and proposed
facilities will play a crucial role.
The high-luminosity upgrade of the LHC (HL-LHC), which will operate between 2027
and the late 2030s, will deliver a total integrated luminosity of more
than $\mathcal{L}=3$ ab$^{-1}$ for ATLAS and CMS and $\mathcal{L}=0.3$ ab$^{-1}$
for LHCb.
This large dataset will provide ample opportunities for PDF studies, in particular
with the measurements
of cross-sections in the few TeV region
for processes such as dijet, top quark pair, direct photon, and
Drell-Yan production,
or the transverse momentum distributions of weak gauge bosons.
These measurements would constrain the large-$x$ behaviour
of the poorly known gluon and sea quarks, which in turn
would lead to improved searches for new heavy particles
predicted in scenarios of new physics beyond the Standard Model~\cite{Beenakker:2015rna}.

%%%%%%%%%%%%%%%%%%%%%%%%%%%%%%%%%%%%%%%%%%%%%%%%%%%%%%%%%%%%%%%%%%%
\begin{figure}[h]
  \begin{center}
    \includegraphics[width=0.99\textwidth]{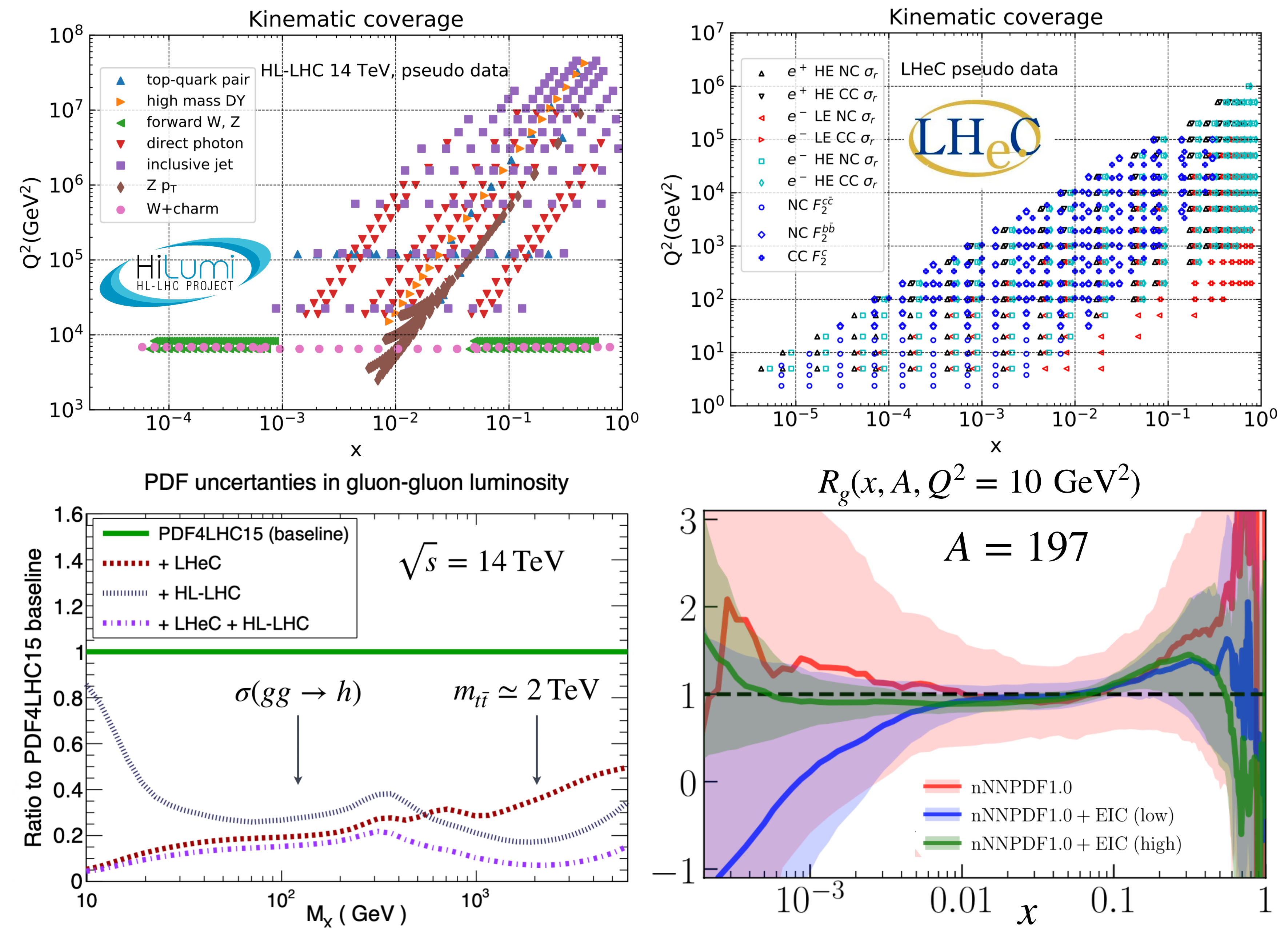}
    \caption{\small Upper plots: the kinematic coverage in the $(x,Q^2)$
      plane that will become available at the HL-LHC (left) and the one that
      would become available if the LHeC is approved (right).
      Bottom left: the reduction in the PDF uncertainties
      of the gluon-gluon luminosity at the LHC (14 TeV)
      once the PDF information contained in the
      LHeC and HL-LHC projections are taken into account,
      compared with the PDF4LHC15 baseline.
      Bottom right: the improvement in the determination
      of the nuclear gluon PDF expected at the EIC, for two
      different scenarios for its center of mass energy.
      \label{fig:future} }
  \end{center}
\end{figure}
%%%%%%%%%%%%%%%%%%%%%%%%%%%%%%%%%%%%%%%%%%%%%%%%%%%%%%%%%%%%%%%%%%%%%%

One proposed future facility that would provide a unique input for both the proton and
nuclear PDFs would be
the Large Hadron electron Collider (LHeC)~\cite{AbelleiraFernandez:2012cc}.
The LHeC would extend the kinematical
coverage achieved at HERA by more than one order of magnitude
at small-$x$ and at large-$Q^2$,
while operating both with proton and with light and
heavy nuclear beams.
The upper panels of Fig.~\ref{fig:future}
display the kinematic coverage in the $(x,Q^2)$
plane of the HL-LHC and the LHeC.
This comparison highlights their complementary,
with the LHeC providing a superior handle in the small-$x$ region
(and in particular allowing for tests of novel QCD dynamics)
with the HL-LHC offering unparalleled reach in the high-energy frontier.
The bottom left panel of Fig.~\ref{fig:future}
displays the expected reduction in the PDF uncertainties
of the gluon-gluon luminosity at the LHC (14 TeV) based
on the LHeC and HL-LHC pseudo-data projections
presented in~\cite{Khalek:2018mdn,AbdulKhalek:2019mps}.
In these forecasts, the baseline is taken to be
the PDF4LHC15 set~\cite{Butterworth:2015oua}, which is  represents our current
best knowledge of the proton PDFs.
One can observe that
in the most favorable scenario, where both HL-LHC and LHeC
operate simultaneously, one might achieve an uncertainty reduction
by up to an order to magnitude for several values of the final state invariant mass $M_X$.

The  Electron Ion Collider (EIC)~\cite{Accardi:2012qut}, an US-based proposal, would
also be able to scrutinise the properties of
nucleons and nuclei with unprecedented detail.
The bottom right panel of Fig.~\ref{fig:future}
shows the improvement in the determination
of the nuclear gluon PDF expected at the EIC for two
scenarios for its center-of-mass energy in the case of the nNNPDF1.0
analysis.
These projections indicate that the EIC could pin down
the nuclear gluon modifications down to $x\simeq 10^{-4}$, and would thus
allow one to probe the possible onset of new QCD dynamics such as non-linear
(saturation) effects.
Furthermore, the EIC would have the unique feature
of being able to chart the proton spin structure in the small-$x$ regime,
as well as its three-dimensional structure for the first
time in a wide kinematic range.

As mentioned before,
another possible direction in which one should expect future progress in the field of
nucleon structure to be made is in the combination of individual QCD fits.
Let me illustrate what it is meant by this with an specific example.
Several proton PDF fits include data taken on nuclear targets, however
they neglect to account for the effects of nuclear modifications.
Furthermore, most nuclear PDF fits assume a proton PDF
baseline (and its uncertainties) that has been extracted by other groups with
in general different methodologies and input assumptions.
Such a situation is far from optimal, since the assumption that the
extraction of proton and nuclear PDFs can be decoupled from each other is not
justified anymore given recent progress in experimental data, theory
calculations, and methodological developments.
Indeed, a more robust approach would  extract simultaneously
the free nucleon ($A=1$) and the nuclear ($A > 1$) PDFs from a single
QCD analysis, and thus be able to keep track of their mutual interplay.

Similar considerations apply to other aspects of the QCD fitting
paradigm.
For instance, there is a natural cross-talk between the (un)polarised
PDFs and the hadron fragmentation functions, which are connected
by means of the semi-inclusive DIS (SIDIS) processes.
SIDIS differs from standard DIS because one of the hadrons in the final state
has been identified.
This implies that the cross-section for this process depends both
on the PDFs of the initial state proton as well as on the fragmentation
functions of the final state hadron.
Therefore,  one should strive to determine simultaneously
the (un)polarised PDFs and the fragmentation functions
from the same joint QCD analysis, as done for example in~\cite{Ethier:2017zbq,Sato:2019yez}.
This approach can provide information on the partonic properties
that is not available via other channels, for instance
to constrain the strange content of the proton
from unpolarised SIDIS measurements~\cite{Borsa:2017vwy,Sato:2019yez}.

In the long term, the community
should aim to assemble a truly global analysis of non-perturbative
QCD objects to extract simultaneously the unpolarised and polarised
proton PDFs, the nuclear PDFs, and the hadron fragmentation
functions.
However, there are two main requirements
that need to be satisfied in order to be able to carry out such an ambitious goal.
The first one is 
methodological progress in the fitting side that ensures that the large parameter space that arises
once all QCD objects are jointly extracted can be explored efficiently.
The second is the availability of 
new facilities that can provide suitable experimental measurements to constrain
all the relevant non-perturbative QCD objects and their correlations.
In this context, a machine such as the Electron Ion collider would be particularly suited
for the realisation of this ultimate ``integrated'' global QCD analysis.

To summarise, the investigation of the  partonic content of nucleons and nuclei
represents a thriving research field
which allows us to tackle pressing open questions in particle, nuclear, and astro-particle
physics.
As discussed here, several breakthroughs in the recent years have made possible a greatly improved
understanding of the quark and gluon substructure of nucleons and nuclei.
While several challenges lie ahead, one can be confident that ongoing progress from
theory, methodology, and experiment will make possible bypassing them and pushing
forward the frontier of our studies of the quark
and gluon parton distributions.

\paragraph{Acknowledgments.}
J.~R. has been supported by the European Research Council Starting
Grant ``PDF4BSM'' as well as by the  Netherlands Organization for Scientific
Research (NWO).

\bibliography{oxford}

\end{document}